\documentstyle[aps,graphicx,twocolumn,cite,floats]{revtex}
\begin{document}           
\draft
\wideabs{
\title{Excitons in a Photosynthetic Light-Harvesting System: A
Combined Molecular Dynamics/Quantum Chemistry and Polaron Model Study}
\author{Ana Damjanovi\'{c}, Ioan Kosztin, and Klaus Schulten}
\address{Beckman Institute and Department of Physics, University of
  Illinois, Urbana, IL 61801, USA}
\date{\today} 
\maketitle
\begin{abstract}
  The dynamics of pigment-pigment and pigment-protein interactions in
  light-harvesting complexes is studied with a novel approach that
  combines molecular dynamics simulations with quantum chemistry
  calculations and a polaron model analysis. The molecular dynamics
  simulation of light-harvesting complexes was performed on an 87,055
  atom system comprised of an LH-II complex of {\em Rhodospirillum
    molischianum} embedded in a lipid bilayer and surrounded with
  appropriate water layers. The simulation provided information about
  the extent and timescales of geometrical deformations of pigment and
  protein residues at physiological temperatures, revealing also a
  pathway of a water molecule into the B800 binding site, as well as
  increased dimerization within the B850 BChl ring, as compared to the
  dimerization found for the crystal structure. For each of the 16 B850
  BChls we performed 400 {\em ab initio} quantum chemistry calculations
  on geometries that emerged from the molecular dynamical simulations,
  determining the fluctuations of pigment excitation energies as a
  function of time. From the results of these calculations we construct
  a time-dependent Hamiltonian of the B850 exciton system from which we
  determine within linear response theory the absorption spectrum.
  Finally, a polaron model is introduced to describe both the excitonic
  and coupled phonon degrees of freedom by quantum mechanics. The
  exciton-phonon coupling that enters into the polaron model, and the
  corresponding phonon spectral function are derived from the molecular
  dynamics and quantum chemistry simulations.  The model predicts that
  excitons in the B850 bacteriochlorophyll ring are delocalized over
  five pigments at room temperature. Also, the polaron model permits the
  calculation of the absorption spectrum of the B850 excitons from the
  sole knowledge of the autocorrelation function of the excitation
  energies of individual BChls, which is readily available from the
  combined molecular dynamics and quantum chemistry simulations. The
  obtained result is found to be in good agreement with the
  experimentally measured absorption spectrum.
\end{abstract}

}

\section{Introduction}

Life on earth is sustained through photosynthesis. The initial
step of photosynthesis involves absorption of light by the
so-called light-harvesting antennae complexes, and funneling of
the resulting electronic excitation to the photosynthetic reaction
center~\cite{VANG94,HU97B,HU98,SUND99}.  In case of purple
bacteria, light-harvesting complexes are ring-shaped aggregates of
proteins that contain two types of pigments, bacteriochlorophylls
(BChls) and carotenoids.

\begin{figure}
\begin{center}
\includegraphics[clip,width=3.in]{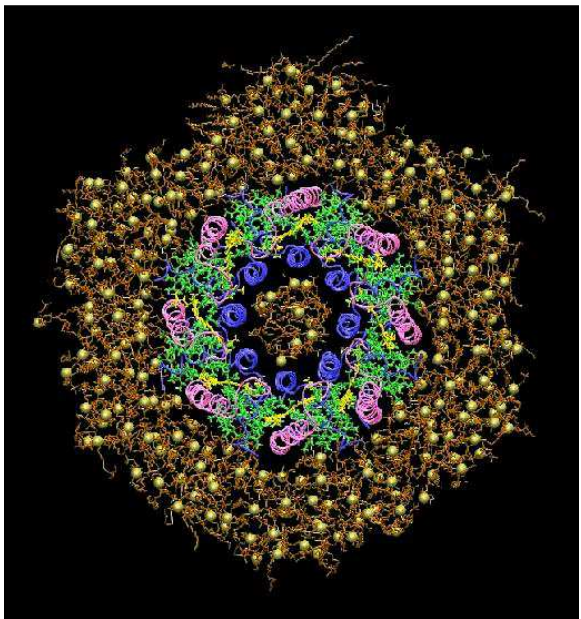}\\
\includegraphics[clip,width=3.in]{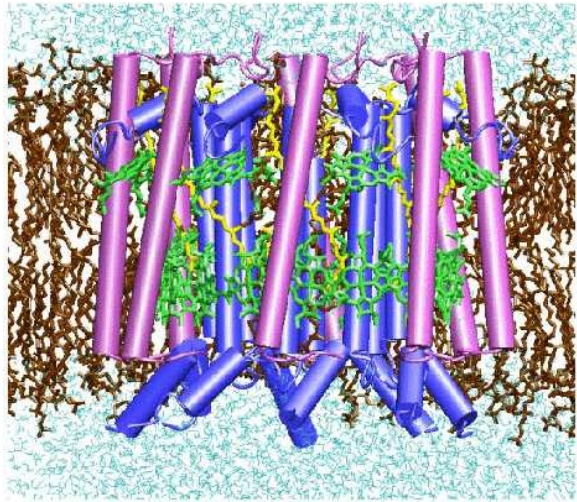}\\[1ex]
\caption{\textbf{Top:} Top view of the LH-II octameric complex of {\it
    Rs.\,molischianum}~\protect\cite{KOEP96} embedded into a lipid
  bilayer that fits into a hexagonal unit cell. The simulated system
  consists of an infinite, periodic repeat of this unit cell.  The
  lipids, 267 molecules of palmitoyloleoylphosphatidylcholine (POPC),
  are shown in brown, with phosphorus atoms in yellow. The $\alpha$- and
  $\beta$- protein subunits of LH-II are shown in blue and magenta,
  respectively; bacteriochlorophylls are shown in green, carotenoids in
  yellow.  \textbf{Bottom:} Side view of LH-II in the lipid-water
  environment. For clarity, the front half of the lipids is not
  displayed. The color code is the same as for the top view; however,
  the $\alpha$-helical segments of the protein subunits are rendered as
  cylinders.  Water molecules are shown in light-blue. One can recognize
  two ring-shaped clusters of BChl molecules (green). A ring of eight
  (only four visible) BChls at the top is formed by the B800 BChls; a
  ring of sixteen BChls (only partially visible) at the bottom is formed
  by 16 B850 BChls. (Produced with the program VMD~\cite{HUMP96}).}
\label{fig:dislh2}
\end{center}
\end{figure}

\begin{figure}
\begin{center}
\includegraphics[clip,width=3.4in]{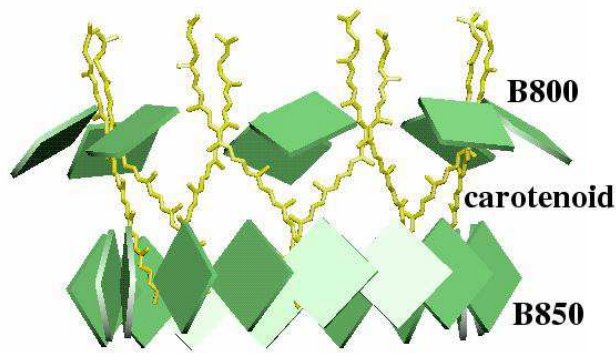}
\caption{Side view of the arrangement of BChl and carotenoid pigments in
  LH-II with the protein components of LH-II stripped away. BChls are
  presented as green squares. Sixteen B850 BChls are arranged in the
  bottom ring and eight B800 BChls in the top ring. Eight carotenoids,
  shown in yellow, were identified in the crystal structure at the
  positions shown (produced with the program
  VMD~\protect\cite{HUMP96}).}
\label{fig:structure}
\end{center}
\end{figure}

Recent solution of the atomic level structure of the light-harvesting
complex II (LH-II) of two species of purple bacteria, {\em
Rhodopseudomonas (Rps.) acidophila}~\cite{MCDE95} and {\em
Rhodospirillum (Rs.) molischianum}~\cite{KOEP96}, has opened the door
to an understanding of the structure-function relationship in these
proteins. Fig.~\ref{fig:dislh2} shows the structure of LH-II of {\em
Rs.~molischianum} in its membrane-water environment. LH-II is an
octamer, consisting of eight $\alpha$- and eight $\beta$- protein
subunits which bind non-covalently the following pigments: eight BChls
that absorb at 800~nm and are referred to as B800 BChls, 16 BChls that
absorb at 850~nm, and are referred to as B850 BChls, and eight
carotenoids which absorb at 500~nm. The pigment system of LH-II is
shown in Fig.~\ref{fig:structure}.  The difference in absorption
maxima between various pigments funnels electronic excitation within
LH-II; pigments with higher excitation energy pass on their excitation
to the pigments with lower excitation energy. As a result, energy
absorbed by the carotenoids, or the B800 BChls, is funneled into the
B850 ring~\cite{HU97}.  The B850 ring, in turn, transfers the
electronic excitation to a BChl ring in another light-harvesting
complex, LH-I, which is directly surrounding the photosynthetic
reaction center. The B850 ring, thus, serves as a link in a chain of
excitation transfer steps which result ultimately in excitation of the
photosynthetic reaction center~\cite{HU97}.

The B800 and B850 BChls have in common a tetrapyrol moiety that
includes the electronically excited $\pi$-electron system; they
differ only in the length of their phytyl chain~\cite{KOEP96}. The
difference in the absorption maxima arises through the difference
in their ligation sites. Protein residues at the ligation site,
especially aromatic, charged or polar ones, as well as hydrogen bonds between pigment and protein residues can shift the
absorption maxima of pigments~\cite{GALL97}. However, the
difference in absorption maxima is also evoked through excitonic
interactions between the B850 BChls~\cite{STUR96}. The
$\pi$-electron systems of neighboring B850 BChls are in van der
Waals contact~\cite{KOEP96}, giving rise to strong couplings
between their electronic excitations. This coupling leads to
coherently delocalized electronic excitations, so-called
excitons~\cite{DAVY62}. 

What is intriguing about the light-harvesting complexes of purple
bacteria is their circular symmetry: LH-II of {\em
Rs.~molischianum} is an octamer;  LH-II of {\em Rps.~acidophila}
is a nonamer~\cite{MCDE95}; electron density projection maps
indicate that LH-I of {\em Rs.~rubrum} is a
hexadecamer~\cite{KARR95}. This poses the question whether the
circular shape plays a functional role. In~\cite{HU97} we have
constructed an effective Hamiltonian to describe the excitons
formed by the Q$_y$ excitations of B850 BChls. The Hamiltonian was
assumed to exhibit a perfect 8-fold symmetry axis for a ring of 16
B850 BChls.  In the case of such ideal symmetry, stationary
states, the excitons, have optical properties characteristic of
the symmetry.  The dipole strength is unequally distributed among
the excitonic states.  In fact, two degenerate excitonic states
carry almost all of the dipole strength. This property facilitates
overall absorption, since the two degenerate  states
each absorb with the strength of eight BChls. Furthermore, an
energy trap is formed; the lowest exciton state carries very small
dipole strength, making fluorescence negligible. As a result, this
state, when populated through thermal redistribution, traps
effectively the excitation energy.  However, this scenario holds
only as long as the symmetry is not broken significantly.

Interactions of BChls with its protein environment in LH-II of
{\sl Rs.\ molischianum}, together with thermal motions, give rise
to static and dynamic disorder which manifests itself through
distortions of the 8-fold symmetry. The classification of disorder
is based on the time-scale of protein thermal motions compared to
the characteristic times of absorption, fluorescence, and
excitation transfer. If the protein motion is slow compared to the
latter processes, the effect of the environment can be considered
as static; otherwise, the effect is dynamic. The nature of the
disorder (static or dynamic) is reflected in the line-shape
functions characterizing the energy dependence of optical
absorption and emission processes. It is important to know these
functions when studying energy transfer between two molecules
since their overlap determines the transfer rate.

The disorder in LH-II affects the energy level structure and
optical properties of excitonic states, that in turn influence
absorption, fluorescence and excitation transfer.  To understand
the dynamics of LH-II it is necessary to quantify the extent of
disorder. One measure of the effect of disorder is the exciton
delocalization length (see for example~\cite{MEIE97,RAY99}). For a
completely symmetric aggregate at temperature T = 0 K, the exciton
delocalization length is equal to the total number of BChls, in
the case of LH-II of {\em Rs.~molischianum}, sixteen. With
increased disorder the coherence of the delocalized states is
lost, and the excitons become confined to a smaller number of
BChls~\cite{RAY99}.

Exciton delocalization in light-harvesting complexes of purple
bacteria has been a subject of extensive experimental study.
Non-linear absorption~\cite{LEUP96}, circular dichroism~\cite{KOOL97},
femtosecond transient absorption~\cite{KENN97},
fluorescence~\cite{MONS97}, pump-probe~\cite{PULL96}, photon
echo~\cite{YANG2001} and single-molecule experiments~\cite{VANO99},
techniques have been employed to study exciton delocalization length,
resulting in vastly different estimates which range from
delocalization over two BChls~\cite{JIME97}, to delocalization over
the entire ring~\cite{LEUP96}. One of the obvious explanations for
different estimates of exciton delocalization are differences in the
definition of observables as well as differences in the temperature at
which experiments were performed.

It appears desirable to complement the experimental observations
of the effect of disorder on the exciton system of light
harvesting proteins by simulations.  Such simulations can provide
a detailed, atomic level picture of the disorder and its effect.
However, the required simulations are extremely demanding.  On the
one hand, one needs to simulate the classical motion of light
harvesting proteins in their natural lipid-water environment under
periodic boundary conditions (to avoid finite size effects) and
under the influence of full electrostatics (dipolar forces play a
key role in membrane systems).  On the other hand, one needs to
carry out, in combination with the classical simulations, quantum
chemical calculations that account for the coupling between
protein dynamics and electronic properties of the BChl systems.
Fortunately, the necessary calculations have become feasible
recently. Naturally, limitations in regard to short sampling times
and still rather crude quantum chemical descriptions still exist.
Nevertheless, a description of static and dynamic disorder based
on computer simulations, rather than on {\sl ad hoc} assumptions,
would constitute a great step forward in our understanding of the
light harvesting apparatus in photosynthesis.  This is
particularly true since the apparatus is based on aggregates of
numerous pigments, the functional properties of which are
sensitive to thermal fluctuations. A model that includes complete
information on the actual disorder and its effect on the
electronic properties provides also an opportunity to improve the
analysis of experimental data.

In this paper we investigate the effect of thermal disorder on excitons
in LH-II at room temperature. We employ a combined molecular
dynamics/quantum chemistry approach to establish an effective
time-dependent Hamiltonian for the Q$_y$ excitations of B850 BChls. This
Hamiltonian is used in a quantum mechanical treatment from which the
spectral characteristics of the B850 BChl system results. The
description is then recast into a polaron model that treats both nuclear
motion and electronic degrees of freedom on an equal (quantum
mechanical) footing.  The theory underlying our analysis of spectral
properties is outlined in chapter II; the methods employed for combined
molecular dynamics/quantum chemistry simulations are described in
chapter III; results, together with the theory of the polaron model, are
presented and discussed in chapters IV and V. Conclusions are stated in
chapter VI.

\section{Theory and Computation of Spectral Properties}

We describe the electronic properties of the B850 BChls in LH-II
under the assumption that the rest of LH-II, i.e., protein, B800
BChls, and lycopenes, play merely the role of a thermal bath. The
total Hamiltonian for this BChl-bath system can be written
\begin{equation}
\label{eq:ddis_1}
\hat H \; = \; \hat H_{BChl} \; + \; \hat H_{BChl-Bath} \; + \; \hat H_{Bath} \, .
\end{equation}
The individual Hamiltonians in Eq.~(\ref{eq:ddis_1}) can be further
decomposed as
\begin{eqnarray}
&&\hat H_{BChl} \; = \; T_{el} \; + \; T_{nuc} \; + \; V_{el-el}
\; + \; V_{el-nuc} \; + \; V_{nuc-nuc} \label{eq:ddis_2} \\ &&\hat
H_{BChl-Bath} \;  =  \; V_{el-Bath} \; + \; V_{nuc-Bath}
\label{eq:ddis_3} \\ &&\hat H_{Bath} \;  =  \;  T_{Bath} \; + \;
V_{Bath-Bath}\; . \label{eq:ddis_4}
\end{eqnarray}
The indices of the quantities introduced in
Eqs.(\ref{eq:ddis_1}-\ref{eq:ddis_4}) are chosen in an obvious
manner to define the partial Hamiltonians sufficiently.   Due to
the large mass difference of electrons and nuclei, we separate
BChl degrees of freedom into ``light'' and ``heavy''ones,  i.e.,
electronic and nuclear. We introduce the following set of indices;
$r = (\vec{r}_1, \vec{r}_2, \cdots, \vec{r}_{N_{el}})$, for all of
the electronic degrees of freedom of BChl; $R = (\vec{R}_1,
\vec{R}_2, \cdots, \vec{R}_{N_{nuc}})$, for all of the nuclear
degrees of freedom of BChl; and $Z = (\vec{Z}_1, \vec{Z}_2,
\cdots, \vec{Z}_{N_{Bath}})$, for all of the bath degrees of
freedom. Index $Z$ labels bath atoms as a whole, i.e., the
splitting of the bath degrees of freedom into nuclear and
electronic is neglected.

According to the Born-Oppenheimer (BO) approximation~\cite{MAY2000} we
assume that the electrons move adiabatically in a potential that
depends parametrically on the atomic coordinates of BChl and bath. If
$\hat H_{BChl}$ describes a single BChl, the above approximation
applies well. The energy gap between ground state and lowest singlet
excited state Q$_y$, the relevant states, measures about 1.5~eV, which
is too large for vibrationally induced transitions to occur, i.e., the
BO approximation holds well.  However, if $\hat H_{BChl}$ describes an
entire B850 BChl ring (comprised of 16 BChls), energy gaps between 16
excitonic states are much smaller and vibrations might induce
transitions between these states. Therefore, we will assume that $\hat
H_{BChl}$ describes only one of the B850 BChl, while the remaining
B850 BChls will join the bath system.  From this description we will
derive the characteristics that enter into a so-called polaron model
which describes vibronic coupling to the electronic degrees of
freedom, treating both excitons and vibrations quantum mechanically.

Within the framework of the BO approximation we construct the
Schr\"odinger equation for the electronic wavefunction, which depends
parametrically on the nuclear coordinates $R$ and coordinates of bath
atoms $Z$~\cite{MAY2000}
\begin{equation}
\label{eq:ddis_6}
\hat H^i_{el}(R,Z) \, \phi^i_a(r_i;R,Z) \; = \; \epsilon^i_a(r_i;R,Z) \, \phi^i_a(r_i;R,Z)  \; .
\end{equation}
Here $\hat H^i_{el}$ is the electronic Hamiltonian of the $i$-th B850
BChl
\begin{equation}
\label{eq:ddis_6a}
\hat H^i_{el}(R,Z) \; = \; T_{el} \; + \; V_{el-el} \; + \; V_{el-nuc}  \; + \; V_{el-Bath} \; ,
\end{equation}
$\epsilon^i_a$ and $\phi^i_a$ are the $a$-th eigenvalue and
eigenvector, respectively, of the $i$-th BChl. At the moment we are
only interested in the ground and the $Q_y$ electronic state of
BChl. The electronic problem stated in
Eqs.~(\ref{eq:ddis_6},\ref{eq:ddis_6a}) has been solved with the
quantum chemistry package {\em Gaussian 98}~\cite{GAUS98}, as
explained in section III.

The equations of motions of the nuclear and bath degrees of freedom
are solved in the framework of the BO approximation, i.e., neglecting
the non-adiabatic coupling~\cite{MAY2000,BAYM73}
\begin{eqnarray}
\label{eq:ddis_7}
&&\hat H_{nuc+Bath}(R,Z) = T_{nuc} \; + \; T_{Bath} \, + \; V_{nuc-nuc}(R)\\
&&+  \; V_{Bath-Bath}(Z) + V_{nuc-Bath}(R,Z) + \epsilon^a_i(R,Z)\;. \nonumber
\end{eqnarray}

Dynamics of the nuclear and bath degrees of freedom can be described
through classical molecular dynamics simulations.  The trajectory of
nuclear and bath coordinates ($R(t_k),Z(t_k)$) for various simulation
times $t_k$ can be generated, as described also in section III. By
treating $R(t_k)$ and $Z(t_k)$ as parameters, energies of the ground
state, $\epsilon^i_0(t_k)$, and of the excited Q$_y$ state,
$\epsilon^i_{Q_y}(t_k)$, will be calculated for each BChl $i$,
$i=1,2,\ldots,16$.

\subsection{Time-dependent Hamiltonian of the B850 Ring}

We construct now the time-dependent electronic Hamiltonian $\hat H(t)$
of the BChl ring. The time-dependence of $\hat H(t) = \hat
H(R(t),Z(t))$ arises through vibrational motions of the nuclear and
bath degrees of freedom. The electronic Hamiltonian is expressed in
the basis comprised of the ground state of the B850 ring
\begin{equation}
\label{eq:ddis_8}
\vert 0 \rangle \; = \; \prod_i \vert \phi^i_0 \rangle
\end{equation}
and of the sixteen Q$_y$ excitations of individual BChls
\begin{equation}
\label{eq:ddis_9}
\vert i \rangle \; = \; \vert \phi^i_{Q_y} \rangle \prod_{j \neq i} \vert \phi^j_0 \rangle \; ,
\end{equation}
where $\vert \phi^i_0 \rangle$ and $\vert \phi^i_{Q_y} \rangle$ denote
the ground and the $Q_y$ excited state of the $i$th BChl.  It is assumed
that the states defined in (\ref{eq:ddis_8}, \ref{eq:ddis_9}) are
time-independent and orthonormal, i.e., $\langle i \vert j \rangle \, =
\, \delta_{i,j}$ for $i, j \, = \, 0, 1, \ldots$. The Hamiltonian now
reads~\cite{MAY2000}
\begin{equation}
\label{eq:ddis_10}
\hat H(t) \; = \; H_0(t) \vert 0 \rangle \langle 0 \vert \; + \; \hat H_1(t) \sum_{i=1}^{16} \vert i \rangle \langle i \vert\; .
\end{equation}
where
\begin{equation}
\label{eq:ddis_11}
H_0(t) \; = \; \sum_i \epsilon^i_0(t)
\end{equation}
and
\begin{eqnarray}
\label{eq:ddis_12} 
\hat H_1(t) = H_0(t)  &+&
\sum_{i,j=1}^{16} [\, \delta_{i,j} (\epsilon^i_{Q_y}(t) -
\epsilon^i_0(t)) \\
&+& \, (1 \, - \delta_{i,j}) \, W_{i,j}(t) ] \,
\vert i \rangle \langle j \vert\, ]\; .\nonumber
\end{eqnarray}
The right-hand term in Hamiltonian~(\ref{eq:ddis_12}) describes the
interactions between Q$_y$ excitations of all the BChls of the B850
ring, and can be rewritten as
\begin{equation}
\label{eq:ddis_13}
\hat H(t)^{exc} =
\pmatrix{
\epsilon_1(t)&&&&&&\cr
&\epsilon_2(t)&&&&&\cr
&&.&&W_{ij}(t)&&\cr
&&&.&&&\cr
&&&&.&&\cr
&&W_{ji}(t)&&&.&\cr
&&&&&&\epsilon_{16}(t)\cr} \; .
\end{equation}
Here $\epsilon_i(t)= {\epsilon^i_{Q_y}(t) - \epsilon^i_0(t)}$ and
$W_{i,j}(t)$ is the induced dipole -- induced dipole coupling
between BChls $i$ and $j$
\begin{equation}
W_{ij}(t) \; = \;  C  \left(\, {\vec{n}_i(t) \cdot  \vec{n}_j(t) \over {r_{ij}(t)}^{3}} - {3 (\vec{r}_{ij}(t) \cdot \vec{n}_i(t)) \;  (\vec{r}_{ij}(t)  \cdot \vec{n}_j(t)) \over {r_{ij}(t)}^{5}} \, \right) \; ,
\label{eq:ddis_14}
\end{equation}
where $\vec n_{j}(t)$ are unit vectors describing the direction of the
transition dipole moments $\vec{d}_j$ of the ground state $\rightarrow$
Q$_y$ state transition of the $j$-th BChl, i.e., $\vec n_{j}(t) = \vec
d_{j}(t)/\vert \vec d_{j}(t) \vert$, and it points from N atom of pyrol
II to the N atom of pyrol IV in BChl $j$ at snapshot $t$. $\vec
r_{jk}(t)$ connects the coordinates of Mg atoms of BChl $j$ and BChl $k$
at time $t$. C is a constant proportional to the square of the magnitude
of the transition dipole, assumed to be constant, and is set to 21.12
\AA$^3 \, eV$. The latter value is obtained when a computationally
determined value of C = 64.38 \AA$^3 \, eV$~\cite{HU97}, which
corresponds to a transition dipole moment of 11~D, is rescaled to match
the experimental value 6.3~D~\cite{VISS91} of the transition dipole
moment. For simplicity we assume that the coupling between neighboring
(and all other) BChls is given by $W_{ij}(t)$, even though the short
distance between neighboring BChls does not justify limitation to only
the leading term of a multipole expansion. We note, however, that the
amplitude of calculated variations of $W_{ij}(t)$ is two orders of
magnitude smaller than that for the diagonal matrix elements
$\epsilon_i(t)$ (see results section), and that the error involved in
the treatment will not alter the qualitative behavior of the system.

The formal solution of the time-dependent Schr\"odinger equation for
Hamiltonian $\hat H(t)^{exc}$~\cite{BAYM73}
\begin{equation}
\label{eq:ddis_15}
i \, {\hbar} \, \frac{\partial \vert \Psi(t)\rangle^{exc}}{\partial t} \; = \; \hat H(t)^{exc} \, \vert \Psi(t) \rangle^{exc} \; ,
\end{equation}
can be written
\begin{equation}
\label{eq:ddis_16} \vert \Psi(t) \rangle^{exc} \; = \; \tilde
U(t,t_0) \, \vert \Psi(t_0)\rangle^{exc} \; ,
\end{equation}
where $\tilde U(t,t_0)$ is the time-development operator. By
dividing the time interval $[t,t_0]$ into $N$ smaller intervals,
$\tilde{U}(t,t_0)$ can be written $(t\, = \, t_N)$
\begin{equation}
\label{eq:ddis_17} \tilde U(t,t_0) \; = \; \tilde U(t_N,t_{N-1})
\, \tilde U(t_{N-1},t_{N-2}) \, \cdots \, \tilde U(t_1,t_0).
\end{equation}
If $N$ is sufficiently large,  $\hat H(t)^{exc}$ can be considered
to be constant in each time interval $[t_k,t_{k+1}]$, i.e., for
 $\hat H(t)^{exc} = \hat
H(t_k)^{exc}$ for $t\, \in \, [t_k,t_{k+1}]$. The time-development
operator $\tilde U(t_{k+1},t_{k})$ can then be determined as
\begin{eqnarray}
\label{eq:ddis_18}
 \tilde U(t_{k+1},t_k) &=& \sum_{m} \exp \left[-{i \over \hbar}
   E_{m}^{(k)} \, (t_{k+1} - t_k)\right]\\ 
&&\times \vert \vert m \rangle^{(k)} \, ^{(k)}\langle m \vert \vert \;
.\nonumber 
\end{eqnarray}
Here $E_m^{(k)}$ and $\vert \vert m \rangle^{(k)}$ are eigenvalues and
eigenvectors of $\hat H(t_k)^{exc}$. The eigenvectors can be expanded
\begin{equation}
\label{eq:ddis_18a} \vert \vert m \rangle^{(k)}  \;= \; \sum_j \,
c_{j,m}^{(k)} \, \vert j \rangle.
\end{equation}
The time development operator $\tilde U(t,t_0)$ can be determined from
combined molecular dynamics/quantum chemistry data. Molecular dynamics
simulation will result in a trajectory of coordinates $R(t_k)$,
$Z(t_k)$. Based on the coordinates $R(t_k)$, the values of
$W_{i,j}(t_k)$ in Hamiltonian~(\ref{eq:ddis_13}) can be calculated
through Eq.~(\ref{eq:ddis_14}). The diagonal matrix elements,
$\epsilon_i(t_k)$, are determined with the quantum chemistry program
{\em Gaussian 98}~\cite{GAUS98}, as described in the methods section.

The Hamiltonian $\hat H(t_k)^{exc}$ can be diagonalized, and the
excitation energies $E_n^{(k)}$, and eigenvectors $\vert \vert m
\rangle^{(k)}$ determined. Calculation of $\tilde U(t,t_0)$ and
$\vert \Psi(t) \rangle^{exc}$ follows straightforwardly from
Eqs.~(\ref{eq:ddis_16},\ref{eq:ddis_17},\ref{eq:ddis_18}).

The time development operator $U(t,t_0)$ for the entire
Hamiltonian~(\ref{eq:ddis_10}), $\hat H(t)$, can be easily
calculated from $\tilde U(t,t_0)$. Exploiting orthogonality
between ground and excited states ($\langle 0 \vert i \rangle \, =
\, 0$), and noting that $H_0(t)$ appearing in
Eqs.~(\ref{eq:ddis_10},\ref{eq:ddis_11},\ref{eq:ddis_12}) is a
c-number, one can express
\begin{eqnarray}
\label{eq:ddis_18b}
U(t,t_0) &=& \exp \left[-{i \over \hbar}
\int_{t_0}^t \, dt^\prime H_0(t^\prime)\right] \\ 
&&\times \left[\vert 0
\rangle \langle 0 \vert \; + \; \sum_{ij} [\tilde U(t,t_0)]_{ij}
\vert i \rangle \langle j \vert \right]\; .\nonumber
\end{eqnarray}
The solution of the time-dependent Schr\"odinger equation
associated with  $\hat H(t)$ is
\begin{equation}
\label{eq:ddis_19}
\vert \Psi(t) \rangle = U(t,t_0) \vert \Psi(t_0) \rangle \; .
\end{equation}

\subsection{Absorption spectrum of a coupled chlorophyll aggregate}

In the following we will derive an expression for the absorption
coefficient for the B850 ring, in the framework of linear response
theory.  We note that within the dipole approximation the total
Hamiltonian $\hat H_{tot}$ describing the BChl system and its
interaction with the radiation field can be chosen
\begin{equation}
\label{eq:abs_4} \hat H_{tot} \; = \; \hat H(t) \; - \; \hat{\vec{\mu}}
\cdot \vec{E}(t) \; .
\end{equation}
Here $\hat H(t)$ is the time-dependent Hamiltonian
(\ref{eq:ddis_10}) of the relevant electronic degrees of freedom
of the BChl system, $\hat \mu$ is the dipole moment operator that
is actually a sum of dipole moment operators of individual BChls
$j$, i.e.,
\begin{equation}
\label{eq:abs_5} \hat{\vec{\mu}} \; = \; \sum_j \hat{\vec{\mu}}_j \; .
\end{equation}
$\vec{E}(t)$ is the electric field of the monochromatic radiation
field, conventionally given in complex form
\begin{equation}
\label{eq:abs_5a} \vec E(t) \; = \; \text{Re}\left(E_0 \vec{u}\,
e^{-i\omega t} \right)
\end{equation}
where the unit vector $\vec{u}$ accounts for the polarization of
the radiation field.

The energy absorbed per unit time due to interaction with the
field is
\begin{equation}
\label{eq:abs_5b} {dW\over dt} \; = \; - \, \langle\;
\hat{\vec{\mu}}(t) \; \rangle \cdot \frac{d\vec{E}}{dt} \; .
\end{equation}
Using (\ref{eq:abs_5a}) we can write
\begin{equation}
\label{eq:abs_5c} {dW\over dt} \; = \;  {i \omega \over 2} \,
E_0\, \vec{u} \cdot \langle\; \hat{\vec{\mu}}(t) \; \rangle
\left(\, e^{-i\omega t} \, - \, e^{i \omega t}\right) \; .
\end{equation}
To calculate the absorption rate we need to determine the dipole
moment  expectation value
\begin{equation}
\label{eq:abs_6} \vec m(t) \; = \; \langle \; U^{-1}_{tot}(t)\,
\hat{\vec{\mu}}\,  U_{tot}(t)\;  \rangle
\end{equation}
where $U_{tot}(t)$ is the propagator associated with the
Hamiltonian (\ref{eq:abs_4}) governed by
\begin{equation}
\label{eq:abs_7} i \, \hbar \, \partial_t \, U_{tot}(t) \;  = \; H_{tot}(t)
 U_{tot}(t)\;,
\end{equation}
and the initial condition
\begin{equation}
\label{eq:abs_7a} U_{tot}(-\infty) \; = \; \openone\;.
\end{equation}
We account for the effect of the radiation field in an
approximate fashion decomposing $ U_{tot}(t)$ in the form
\begin{eqnarray}
\label{eq:abs_7b} U_{tot}(t) \; & = & \;U(t) \, \left[\, \openone
\; + \; \delta U(t)\right]
\\ \label{eq:abs_7c}
\delta U(-\infty)\; & = & \; 0
\end{eqnarray}
where $U(t)$ is defined through (\ref{eq:ddis_18b}) for $t_0\, =
\, -\infty$.  One can readily show that $\delta U(t)$ obeys to
leading order
\begin{eqnarray}
\label{eq:abs_7d} i\, \hbar \, \partial_t \delta U(t) \; & \approx
& \;  - \,  \hat{\vec \mu}(t) \cdot \vec E(t)
\\ \label{eq:abs_7e}
 \hat{\vec \mu}(t)\; & = & \; U^{-1}(t)\,\hat{\vec \mu} \, U(t) \;
 .
\end{eqnarray}
The corresponding solution is
\begin{equation}
\label{eq:abs_7f} \delta U(t)\; = \; {i \over \hbar} \,
\sum_{\beta}\, \int_{-\infty}^t dt^\prime \,
\hat\mu_\beta(t^\prime)\, E_\beta(t^\prime)
\end{equation}
where we introduced the $\beta$-th Cartesian components of
$\hat{\vec \mu}(t), \,  \vec E(t)$.  From this follows, using
(\ref{eq:abs_7b}), again accounting only for terms to leading
order
\begin{equation}
\label{eq:abs_7g} m_\alpha(t)\; \approx \; 
\hat\mu_\alpha(t) \; + \; {i\over \hbar} \, \sum_\beta \,
\int_{-\infty}^t dt^\prime \,\langle \, [\, \hat \mu_\alpha(t), \,
\hat\mu_\beta(t^\prime)]\, \rangle\, E_\beta(t^\prime)\; .
\end{equation}
Conventionally, one writes this in the form
\begin{equation}
\label{eq:abs_7h} m_\alpha(t)\; = \; m_\alpha^{(0)} \; + \;
\sum_\beta \int_{-\infty}^\infty dt^\prime 
\,\chi_{\alpha \beta}(t, t^\prime)\, E_\beta(t^\prime)\; .
\end{equation}
where we defined the so-called susceptibility tensor
\begin{equation}
\label{eq:abs_7k} \chi_{\alpha \beta}(t, t^\prime) \; = \; {i\over
\hbar}\, \Theta(t\, - \, t^\prime)\,\langle \, [\, \hat
\mu_\alpha(t), \, \hat\mu_\beta(t^\prime)]\, \rangle
\end{equation}
introducing the Heavyside function $\Theta(t-t^\prime)$.

In the following we will assume that the light harvesting systems
studied are initially in the electronic ground state $\vert 0
\rangle$, i.e., the susceptibility tensor is in the present case
\begin{equation}
\label{eq:abs_13}
 \chi_{\alpha,\beta}(t,t^\prime) \; = \; {i \over {\hbar}} \,
 \Theta(t-t^\prime) \,
 \langle 0 \vert [ \hat \mu_\alpha(t), \hat \mu_\beta(t^\prime)]
 \vert 0 \rangle  \; .
\end{equation}
Inserting the identity operator
\begin{equation}
\label{eq:abs_15} \openone \; = \; \vert 0 \rangle \langle 0 \vert
\; + \; \sum_{k=1}^{16} \vert k \rangle \langle k \vert
\end{equation}
 into expression~(\ref{eq:abs_13}), after some algebra,  results in
\begin{eqnarray}
\label{eq:abs_17a}
 \chi_{\alpha \beta}(t,t^\prime) &=& {i \over {\hbar}} \,
 \Theta(t-t^\prime) \,
\sum_{k,l=1}^{16} \,  \left( \, {d}_{k\alpha} \, {d}_{l\beta} \,
 \left[\tilde U(t) \tilde U^{-1}(t^\prime)\right]_{kl} \right.\\
&&\left. -d_{k\beta} \, {d}_{l\alpha}\, \left[ \tilde U(t^\prime)\, \tilde
U^{-1}(t)\right]_{kl}\, \right) \; .\nonumber
\end{eqnarray}
We have introduced here the real quantities $d_{k\alpha}\, = \,
\langle 0 \vert \hat\mu_\alpha \vert k \rangle$ which denote the
($\alpha$-th Cartesian component of) the transition dipole moments
of the individual BChls $k\, = \, 1, 2, \ldots \, 16$. Using the
unitarity property $[U^{-1}(t)]_{jk}\, = \, [U(t)]^*_{kj}$ one can
write (\ref{eq:abs_17a})
\begin{eqnarray}
\label{eq:abs_17c}
 \chi_{\alpha \beta}(t,t^\prime) &=& {i \over {\hbar}} \,
 \Theta(t-t^\prime) \,
\sum_{k,l=1}^{16} \,  {d}_{k\alpha} \, {d}_{l\beta} \\
&&\times\left( \,
 \left[\tilde U(t) \tilde U^{-1}(t^\prime)\right]_{kl} \; -
\;\overline{ \left[ \tilde U(t)\, \tilde
U^{-1}(t^\prime)\right]_{kl} }\, \right)\nonumber
\end{eqnarray}
from which follows
\begin{equation}
\label{eq:abs_17d}
 \chi_{\alpha \beta}(t,t^\prime) \; = \; - \,  {2 \over {\hbar}} \,
 \Theta(t-t^\prime) \,
\sum_{k,l=1}^{16} \,  {d}_{k\alpha} \, {d}_{l\beta} \, \text{Im}
 \left[\tilde U(t) \tilde U^{-1}(t^\prime)\right]_{kl} \; .
\end{equation}
Employing the notation in Eq.~(\ref{eq:ddis_16}) one can express
$\tilde U(t,t^\prime) \, = \, \tilde U(t)\, \tilde
U^{-1}(t^\prime)$ and conclude
\begin{equation}
\label{eq:abs_17e}
 \chi_{\alpha \beta}(t,t^\prime) \; = \; - \,  {2 \over {\hbar}} \,
 \Theta(t-t^\prime) \,
\sum_{k,l=1}^{16} \,  {d}_{k\alpha} \, {d}_{l\beta} \, \text{Im}
 \left[\tilde U(t,t^\prime)\right]_{kl} \; .
\end{equation}

One can now express the dipole moment expectation value given by
(\ref{eq:abs_7h}) in terms of (\ref{eq:abs_17e}).  In the
following we are actually considering the ensemble average
$\langle \cdots \rangle_e$ of the dipole moment
\begin{equation}
\label{eq:abs_11d_} M_\alpha(t) \equiv \langle
m_\alpha(t)\rangle_e = \langle m^{(0)}_\alpha 
\rangle_e \; + \; \sum_\beta \, \int_{-\infty}^{\infty} \,
dt^\prime \langle \chi_{\alpha \beta}(t, t^\prime)\rangle_e \,
E_\beta(t^\prime)\;.
\end{equation}
We will assume  that $\langle m^{(0)}_\alpha \rangle_e$ is
time-independent.  In calculating $\langle \chi_{\alpha \beta}(t,
t^\prime)\rangle_e$ we are left, according to (\ref{eq:abs_17e}),
with expressions of the type $\langle {d}_{k\alpha} {d}_{l\beta}
\left[\tilde U(t, t^\prime)\right]_{kl}\rangle_e$. We assume that
${d}_{k\alpha} {d}_{l\beta}$ are approximately constant over the
ensemble and can be replaced by $\overline{{d}_{k\alpha}
{d}_{l\beta}}$.   We also assume that the expressions $\langle
\left[\tilde U(t,t^\prime)\right]_{kl}\rangle_e$ are
translationally invariant in time, i.e., are functions of $t\, -
\, t^\prime$ only.  This permits us to express
\begin{equation}
\label{eq:abs_17e1}
 \chi_{\alpha \beta}(t-t^\prime) \; = \; - \,  {2 \over {\hbar}}
 \;  \Theta(t-t^\prime) \,
\sum_{k,l=1}^{16} \,  \overline{{d}_{k\alpha} \, {d}_{l\beta}}
\,\text{Im} \, \langle [\tilde U(t-t^\prime, 0)]_{kl} \rangle_e
\end{equation}
and also
\begin{equation}
\label{eq:abs_11e} M_\alpha(t) \; = \; \langle m^{(0)}_\alpha
\rangle_e \; + \; \sum_\beta \, \int_{-\infty}^{\infty} \,
dt^\prime  \chi_{\alpha \beta}(t-t^\prime) E_\beta(t^\prime) \; .
\end{equation}

In the following we introduce the  Fourier transform of the
susceptibility tensor,
\begin{equation}
\label{eq:abs_11f} \tilde \chi_{\alpha\beta}(\omega) \; = \;
\int_{-\infty}^\infty d\tau \;  \chi_{\alpha \beta}(\tau) \,
e^{i\omega \tau} \; .
\end{equation}
Since in Eq.~(\ref{eq:abs_11e}) both $M_\alpha(t)$ and $E_\beta(t)$ are
real, it follows that $\chi_{\alpha\beta}(\tau)$ must also be real
[cf.~Eq.(\ref{eq:abs_17d},\ref{eq:abs_17e})], and from
Eq.~(\ref{eq:abs_11f}) this yields the expression

\begin{equation}
\label{eq:abs_11fa} \tilde \chi_{\alpha\beta}(\omega) \; = \;
\tilde \chi^*_{\alpha\beta}(-\omega) \; .
\end{equation}
Combining (\ref{eq:abs_5a}) with (\ref{eq:abs_11e}) and
(\ref{eq:abs_11f}), one obtains
\begin{eqnarray}
\label{eq:abs_11h} M_\alpha(t) \; = \; \langle m^{(0)}_\alpha
\rangle_e  &+&  {1\over 2}\,  \sum_\beta \,
\left[ \tilde \chi_{\alpha \beta}(-\omega)\, e^{i\omega t} \right.\\
&+&\left.
 \tilde \chi_{\alpha \beta}(\omega)\, e^{-i\omega t}\right]\, u_\beta \,
E_0 \; . \nonumber
\end{eqnarray}
We can now determine the absorption rate for radiation by employing this
expression instead of $\langle\hat{\vec{\mu}}(t)\rangle$ in
(\ref{eq:abs_5c}). In doing so we consider the time average (denoted by
an overbar) over a cycle of the radiation field and use
$\overline{\exp(\pm i\omega t)}\, = \, 0$. This yields
\begin{equation}
\label{eq:abs_11f_} \overline{{d W\over dt}}\; = \; {1\over 4} i
\omega E_0^2 \; \sum_{\alpha\beta} u_\alpha u_\beta \, [\, \tilde
\chi_{\alpha\beta}(-\omega)\; - \; \tilde
\chi_{\alpha\beta}(\omega)\, ] \; .
\end{equation}
Using  Eq.~(\ref{eq:abs_11fa}) one can express
\begin{equation}
\label{eq:abs_11g} \overline{{d W\over dt}}\; = \;-  {1\over 4} i
\omega E_0^2 \sum_{\alpha\beta} u_\alpha u_\beta \, [\, \tilde
\chi^*_{\alpha\beta}(\omega)\; - \; \tilde
\chi_{\alpha\beta}(\omega)\, ] \; .
\end{equation}

Expression (\ref{eq:abs_11g}) describes the absorption of
radiation with a fixed polarization $\vec{u}\, = \, (u_1, u_2,
u_3)^T$.  We are actually interested in absorption of light of any
polarization and coming from any direction.  Carrying out the
corresponding average $\langle \cdots \rangle_{rad}$ amounts to
replacing in (\ref{eq:abs_11g}) $u_\alpha u_\beta$ by ${2\over
3}\delta_{\alpha\beta}$.  The respective absorption rate is
described by
\begin{equation}
\label{eq:abs_11h1} \left\langle \overline{{d W\over
dt}}\right\rangle_{rad} \; = \;- {1\over 6} i \omega E_0^2
\sum_\alpha\,  [\, \tilde \chi^*_{\alpha\alpha}(\omega)\; - \;
\tilde \chi_{\alpha\alpha}(\omega)\, ]
\end{equation}
which can also be written
\begin{equation}
\label{eq:abs_11j} \left\langle \overline{{d W\over
dt}}\right\rangle_{rad} \; = \; {\omega\over 3} \, {\rm
tr} \,\text{Im} \,  \tilde \chi(\omega)\;  E_0^2 \; .
\end{equation}
With the definition of the field intensity $I\, = \, {c\over 2
\pi} E_0^2$ one can write this
\begin{equation}
\label{eq:abs_11k} \left\langle \overline{{d W\over
dt}}\right\rangle_{rad} \; = \; \alpha(\omega) I \; .
\end{equation}
where the absorption coefficient is~\cite{MAY2000}
\begin{equation}
\label{eq:abs_11l}  \alpha(\omega) \; = \; {2 \pi \omega \over 3 c}
{\rm tr}\,\text{Im} \tilde \chi(\omega)\; .
\end{equation}
Using Eq.~(\ref{eq:abs_17c}) one obtains then the expression
\begin{eqnarray}
\label{eq:abs_19}
 \alpha(\omega) &=& {4 \pi \omega \over 3 \hbar n c}\;{\rm Re} \, \int_0^{\infty} \, dt \, \exp
 \left[i \omega t\right] \\
&&\times\sum_{k,l=1}^{16} \, \overline{\vec{d}_k \vec{d}_l} \, \left(
\langle  \,  [\tilde U(t,0)]_{kl} \rangle_e \; - \; \langle 
[\tilde U(t,0)]^*_{lk} \rangle_e \right)  \,  \; .\nonumber
\end{eqnarray}
Finally, the line shape function can be expresed in terms of the
absorption coefficient \cite{MAY2000}

\begin{equation}
  \label{eq:abs_19a}
  I(\omega) = \frac{3c}{4\pi^2\omega|d|^2}\, \alpha(\omega)\;,
\end{equation}
where $|d|^2$ is the mean square of the dipole moment of the system.

The absorption coefficient for a single BChl follows from the
above expression ~(\ref{eq:abs_19}) by setting
\begin{equation}
\label{eq:abs_20} \tilde U_{jk}(t_m,t_0) \; = \; \delta_{jk} \;
\prod_{\ell=1}^m \exp \left[ -{i \over \hbar}
\epsilon_j(t_{\ell-1})\, (t_\ell - t_{\ell-1}) \right] \; ,
\end{equation}
where $\epsilon_i(t_\ell)$ is the excitation energy of BChl at time
$t_\ell$ of the simulation.  
By inserting (\ref{eq:abs_20}) into Eq.~(\ref{eq:abs_19}), and using
Eq.~(\ref{eq:abs_19a}) one obtains a familiar expression for the
line-shape function~\cite{WANG85} for a single BChl
\begin{eqnarray}
\label{eq:abs_21}
I(\omega) &=& {1 \over 2 \, \pi} \;\text{Re}\!\int_{-\infty}^{\infty} dt \;
 \exp \left[i \, (\omega - {\epsilon_0 \over \hbar})\, t \right] \\
 &&\times\left\langle \exp \, -{i \over \hbar} \, \int_0^t dt^\prime \, \delta
 \epsilon(t^\prime) \; \right\rangle  \;.\nonumber
\end{eqnarray}
Here the sum in~(\ref{eq:abs_20}) has been replaced by an
integral and $\epsilon_0$ is the average value of $\epsilon(t)$ (we
dropped the index $j$, that labels the BChls); we also defined $\delta
\epsilon(t) \; = \; \epsilon(t) - \epsilon_0$ and $t_0\, = \, 0$.

\section{Computational Methods}

In this section we will describe methodological details of the
molecular dynamics (MD) simulation on LH-II as well as of the quantum
chemistry on BChls.

\subsection{Molecular Dynamics Simulation}

The simulated system, shown in Fig.~\ref{fig:dislh2}, consists of
the LH-II protein surrounded by a lipid bilayer and a 35~\AA~layer
of water molecules. The system has been constructed in four steps,
(i) - (iv).

Step (i): Using~X-PLOR~\cite{BRUN92b} we added hydrogen atoms to
the published structure of an LH-II octamer~\cite{KOEP96}, pdb
entry 1LGH. Hydrogen atoms of BChls and lycopene were constructed
by employing the ``add all hydrogens'' command from the molecular
editor in QUANTA~\cite{QUAN97}. The new structure, containing
explicitly all hydrogen atoms, was subjected to a series of
minimization steps using X-PLOR~\cite{BRUN92b}.

Step (ii:) The modeled LH-II structure was placed into an
elementary cell that was periodically extended.  The cell had the
shape of a hexagonal prism.  Accordingly, we placed LH-II into a
hexagonally shaped patch of lipid bilayer. We chose for the
hexagon a side of length $a = $~60~\AA; the height of the
bilayer was about 42~\AA.  

The lipids employed were palmitoyloleoylphosphatidylcholine
(POPC). Ten lipid molecules were placed within the central opening
of the LH-II (see Fig~\ref{fig:structure}a) and 257 lipids were
placed around the protein.  The number of lipids in the center
were chosen according to the available area in the LH-II center
and to the surface area of equilibrated POPC lipids~\cite{HELL93}.
The chosen number is also justified {\sl a posteriori} since the
central opening of LH-II remained constant in size during the
subsequent equilibration described below.

Step (iii): Two water layers, with 35~{\AA} combined thickness, of the
same hexagonal shape as that of the lipid bilayer ($a =$~60~\AA), were
added to both sides of the system, by removing all water molecule within
2.4~\AA\ distance from heavy atoms of protein or lipids.

Step (iv): Electrostatic forces were calculated using the
Particle-Mesh-Ewald method (PME).  To assure neutrality of the
system, required by this method, we added 16 Cl$^-$ ions to the
simulated system, and placed them not closer than 5~\AA~from the
protein residues at positions of minimal electrostatic energy.
However, due to the high mobility of the ions, the original
positions chosen were largely irrelevant.

Input files for the MD simulation were prepared with the program
XPLOR~\cite{BRUN92b}.  We used the CHARMM22 force
field~\cite{MACK92,MACK98} to describe protein and lipids. For water we
employed the TIP3 model~\cite{JORG83}, omitting internal geometry
constraints~\cite{TEET91,DAGG93,STEI93}.  Ground state partial charges
for the geometry minimized structure of BChl (phytyl tail omitted) were
calculated with the program JAGUAR~\cite{JAGU96}, employing the ESP
method (charge, dipole, quadrupole and octupole moment being the ESP
constraints) and a 6-31G$**$ basis set. The calculated charges are
presented in Table~\ref{tab:BChl_charges}. We note here that we used
fixed ground state charges for BChls throughout the simulation. The
remaining force field parameters for the BChls were chosen as those
presented in~\cite{FOLO92,FOLO95}. Force field parameters for lycopene
were assigned with the program QUANTA~\cite{QUAN97} by exploiting
QUANTA's display parameter file.
\begin{table}[t]
\begin{center}
\begin{tabular}{|lc|lc|lc|}
atom&   charge  & atom&  charge   &atom & charge \cr
\hline              \hline
MG  &   1.02571 & CMB &  -0.19250 & O1A &  -0.58093        \cr
NA  &  -0.33307 & C3B &  -0.41235 & O2A &  -0.45950        \cr
NB  &  -0.72980 & CAB &   0.78053 & CP1 &  -0.11067        \cr
NC  &  -0.28498 & CBB &  -0.35230 & HBH &   0.19015        \cr
ND  &  -0.50505 & OBB &  -0.56500 & HCH &   0.23106        \cr
C1A &  -0.28240 & C2C &   0.41596 & HDH &   0.16850        \cr
CHA &   0.46388 & CMC &  -0.25779 & H2A &   0.18719        \cr
C4D &   0.02422 & C3C &   0.11349 & HAA1,2 &  0.05460      \cr
C1B &   0.48104 & CAC &   0.19597 & H3A  & -0.03004        \cr
CHB &  -0.71796 & CBC &  -0.26871 & HMA1,2,3 &  0.132407   \cr
C4A &   0.39874 & C2D &   0.05822 & HMB1,2,3 &  0.0744133  \cr
C1C &   0.23433 & CMD &  -0.46669 & HBB1,2,3 &  0.0998467  \cr
CHC &  -0.65635 & C3D &  -0.30134 & H2C   & -0.05694       \cr
C4B &   0.55591 & CAD &   0.88725 & HMC1,2,3 &  0.0494933  \cr
C1D &   0.17838 & OBD &  -0.60137 & H3C   & -0.01165       \cr
CHD &  -0.28183 & CBD &  -1.10906 & HAC1,2 & -0.03769      \cr
C4C &  -0.13350 & CGD &   1.11971 & HBC1,2,3 &   0.0503367 \cr
C2A &  -0.30887 & O1D &  -0.64336 & HMD1,2,3 &   0.140277  \cr
CAA &   0.08625 & O2D &  -0.38998 & HBD    & 0.30939       \cr
C3A &   0.48383 & CED &  -0.20137 & HED1,2,3 &   0.123983  \cr
CMA &  -0.53352 & CBA &  -0.60870 & HBA1,2  & 0.16122      \cr
C2B &   0.11499 & CGA &   0.97654 & HP11,12   & 0.168895   \cr
\end{tabular}
\end{center}
\caption{Partial charges for geometry optimized BChl 
(phytyl tail omitted), calculated with the program
JAGUAR~\protect\cite{JAGU96}, with the ESP method (charge, dipole,
quadrupole and octupole moment being the ESP constraints) and the
6-31G$**$ basis set. To hydrogen atoms bonded to the same carbon
we assigned an average charge. Charges for the phytyl
tail were those presented in~\protect\cite{FOLO92,FOLO95}}
\label{tab:BChl_charges}
\end{table}
As pointed out above, the simulated system was first subjected to
energy minimizations in XPLOR~\cite{BRUN92b}. Additional
minimizations, equilibration and MD simulations were performed
using the molecular dynamics program NAMD, version
2~\cite{KALE99}. We used the SHAKE constraint for all hydrogen
atoms. For the dielectric constant we chose the value $\epsilon =
1$; the integration time-step during the equilibration phase of
the simulation was 2~fs. The equilibration was carried out for a
time period of 2~ns in an NpT ensemble mode with a total of 87,055
atoms per elementary cell, at a pressure of 1 atm and a
temperature of 300 K for periodic boundary conditions, calculating
electrostatic interactions in full using the PME method. The
cutoff for Ewald summations was 10~{\AA}. Our choice of a
35~\AA~water layer, and a 20~\AA~layer of lipids between two
neighboring LH-II molecules ensures adequate separation between
the proteins. The choice of a hexagonal unit cell, as compared to
a square shaped unit cell, significantly reduced the size of the
simulated system.

After the equilibration phase, we performed two molecular dynamics 
runs, using a 1~fs integration time-step. The first run was performed
for 10~ps, by recording configurations every 100~fs. The second run was
performed for 800~fs, with snapshots of the trajectory recorded every
2~fs. The structures associated with the latter 2~fs snapshots were used
in quantum chemistry calculations of BChl's excitation energies. Due to
a need to perform quantum chemistry calculations for all 16 BChls of the
B850 system, and due to the high computational effort (calculations on
four processors of Silicon Graphics Origin 2000 required for every
snapshot one hour computing time per BChl), we could only perform
calculations for 400 snapshots.

\subsection{Quantum Chemistry Calculations}

Calculations of the ground and excited state energies of individual
BChls have been performed using the program {\em
Gaussian~98}~\cite{GAUS98}. The coordinates and charges of the bath
atoms provided a background charge distribution. The calculations were
performed at the Hartree-Fock/configuration interaction singles
(HF/CIS) level, with the STO-3G basis set. We used this basis set
because it is computationally the least expensive, and because our
test calculations showed that the more sophisticated 6-31G* basis set
resulted in only slightly different fluctuations of excitation
energies. To further speed up calculations, we restricted the active
space for the CIS calculations to the ten highest occupied molecular
orbitals and to the ten lowest unoccupied molecular orbitals, as
suggested in~\cite{MERC99}.

The use of the {\em ab initio} (HF/CIS) method to calculate
excitation energies versus semi-empirical, or classical
calculations was discussed by the authors in~\cite{MERC99}. Even
though the {\em ab initio} method predicted incorrectly the
average excitation energies of BChls (as opposed to the
semi-empirical method), it predicted well the width of the
absorption spectrum, which is determined by the fluctuations of
the energy values around the average value. We, thus, chose to use
the {\em ab initio} method to determine the fluctuations, while
for the average excitation energy we used a value of 1.57~eV.

\section{Simulated Protein Dynamics and Spectra}

Below we will first compare the crystal structure of LH-II of {\em
  Rs.~molischianum}~\cite{KOEP96}, used to initiate our simulations,
with the structure emerging after 2~ns MD equilibration. We will then
present the spectral properties predicted for the B850 system of LH-II
under the influence of thermal fluctuations as described in combined
quantum chemistry, quantum mechanics, and classical mechanics
calculations.  In the following section, we will cast our results into a
polaron model that describes the vibronic coupling of the B850 exciton
system.

\subsection{Structure of LH-II at room temperature}

In order to analyze the geometry of LH-II equilibrated in our MD
simulations we performed a 10~ps MD simulation run.  This permits
us to compare the crystallographically determined structure of
LH-II of {\em Rs.~molischianum} as reported in~\cite{KOEP96} with
the simulated structure.

\subsubsection{B850 BChls}

\begin{figure}[t]
\begin{center}
\includegraphics[clip,width=3.4in]{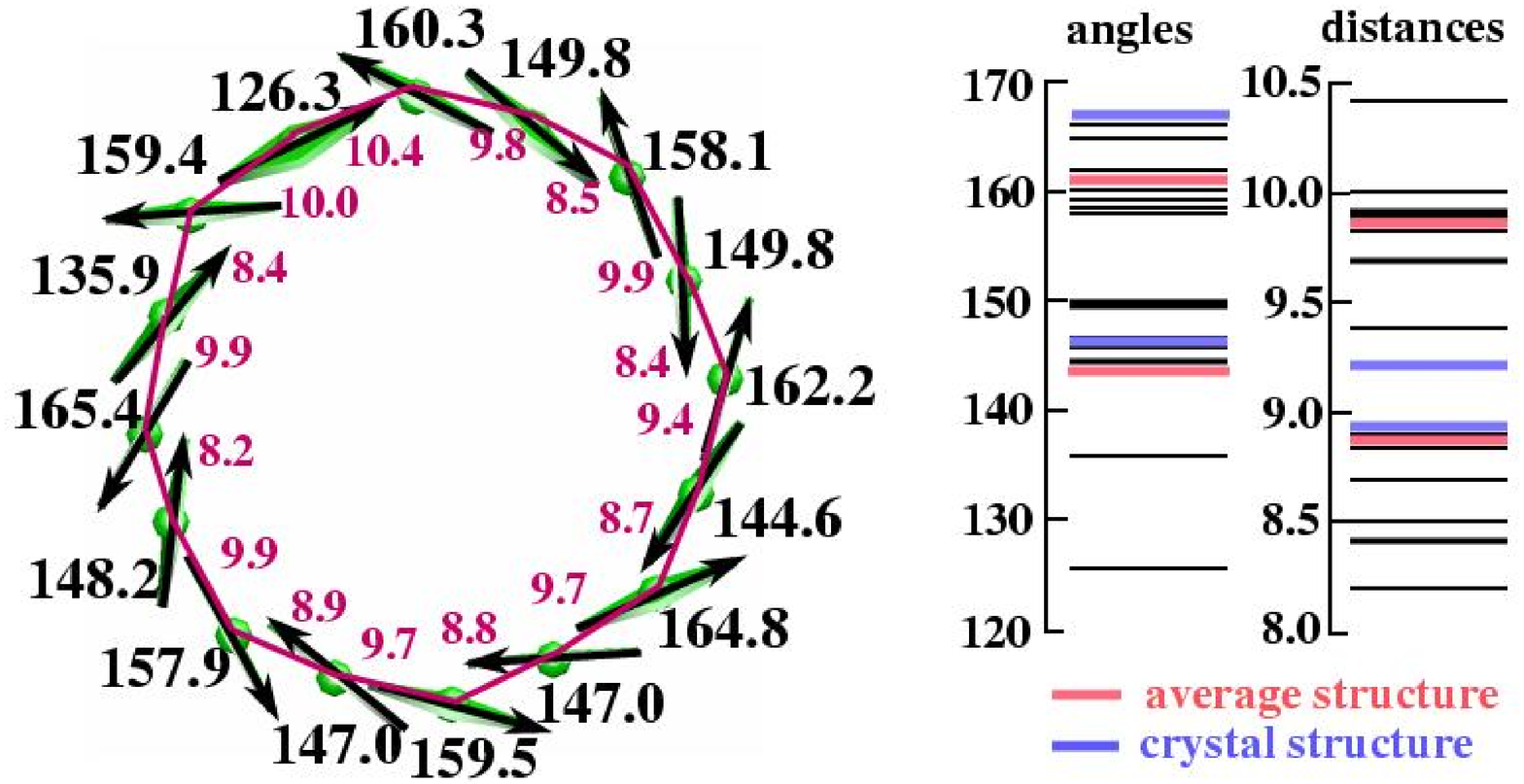}
\caption{
Orientation of B850 BChls resulting from a snapshot of a 10~ps long
molecular dynamics simulation. Orientation of the Q$_y$ transition
dipole moments is indicated by arrows. Angles and distances between
the nearest neighbors (in degrees and in~\AA), averaged over a 10~ps
long simulation, are given on the outside and on the inside of the
ring, respectively.  The same numbers are shown in the right-hand side
of the Figure, and compared to the angles and distances of the
crystal structure.}
\label{fig:mg}
\end{center}
\end{figure}

The Mg--Mg distances between neighboring B850 BChls, averaged over the
10 ps simulation, are shown in Fig.~\ref{fig:mg}. The average Mg--Mg
distances within and between $\alpha\beta$-heterodimers are 9.8~\AA~ and
8.8~\AA, respectively.  They differ from the corresponding averages in
the crystal structure~\cite{KOEP96}, which are 9.2~\AA~within an
$\alpha\beta$-heterodimer and 8.9~\AA~between heterodimers.  Average
angles between neighboring BChls are also shown in Figure~\ref{fig:mg}.
The average angles between neighboring BChls are 160.9$^{\circ}$ within
a $\alpha\beta$-heterodimer, and 143.5$^{\circ}$ between heterodimers.
The corresponding angles in the crystal structure are 167.5$^{\circ}$
and 147.5$^{\circ}$, respectively~\cite{KOEP96}.

We note here that the average angles and distances result from a
10~ps sampling. The possibility that, for a 10~ps long sampling, a
structural inhomogeneity arises can be inferred from the fact that
one of the BChls Mg--Mg inter-heterodimer distance was found to be
much larger (10.4~\AA) than the average distance ($8.6$~\AA). This
structural inhomogeneity will likely disappear for longer sampling
times.

\subsubsection{B800 BChls}
\begin{figure}[t]
\begin{center}
\includegraphics[clip,width=3.4in]{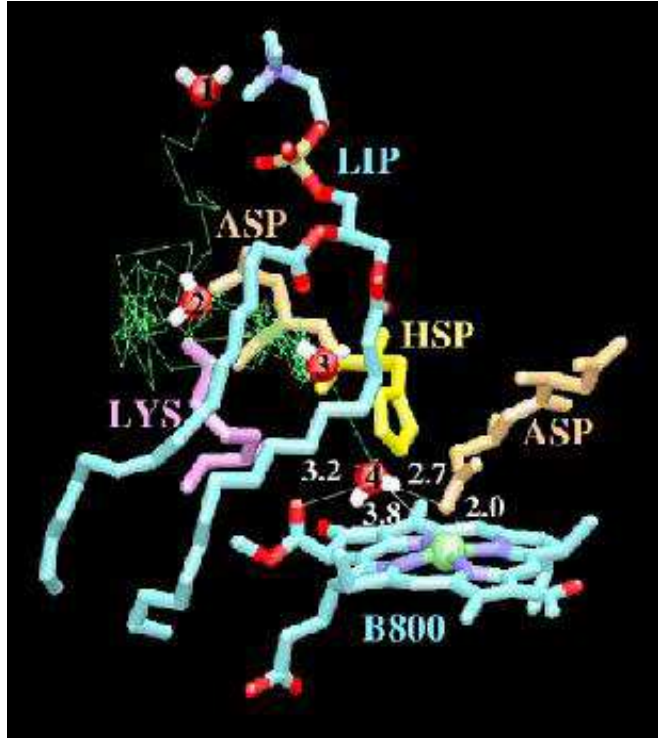}\\[2ex]
\caption{Binding pocket for B800 BChl and surrounding protein and
lipid residues. Shown, in green, is a trajectory of a water
molecule, obtained from a 440~ps long simulation. Snapshots of the
water molecule at times  t = 0 (1),  t = 230~ps (2),  t = 360~ps
(3), t = 362~ps (4) are represented. Distances, averaged over a
10~ps simulation, of the O atom of the water molecule, in position
(4), with key protein and BChl atoms are given.} \label{fig:B800}
\end{center}
\end{figure}

The crystal structure of LH-II of {\em Rs.~molischianum} revealed an
unusual ligation of the B800 BChls~\cite{KOEP96}. In variance with the
customary ligation of BChl's central Mg atom to a His residue (as found
for the B850 BChls), the Mg atom of B800 BChls is coordinated to the
O$\delta_1$ atom of an Asp residue. In addition, an H$_2$O molecule is
found in the vicinity of the ligation site. The initial, minimized
structure of LH-II embedded into a lipid-water environment did not
contain any water molecules within the B800 binding pocket. During the
course of the equilibration, a water molecule diffused into the B800
binding pocket for seven out of eight of the B800 BChls. In
Fig.~\ref{fig:B800} we show a diffusion pathway of a water molecule into
the binding site of one of the B800 BChls, recorded over the first
440~ps of the equilibration run. Representative distances within the
binding pocket, averaged over the 10~ps long simulation, are shown in
Fig.~\ref{fig:B800}: 2.72~\AA~(O atom of H$_2$O to O$\delta 1$ atom of
Asp6), 3.76~\AA~(O atom of H$_2$O to Mg atom of BChl),
3.16~\AA~(carbonyl O2 atom of BChl to O atom of H$_2$O), 2.00~\AA~(Mg
atom of BChl to O$\delta 1$ atom of Asp6). The respective distances
determined for the crystal structure of LH-II are 2.74~\AA, 4.24~\AA,
3.10~\AA~and 2.45~\AA, respectively~\cite{KOEP96}.

\subsection{Time-series analysis of simulation data}

The time series analysis presented in this section is based on an 800~fs
long MD simulation.  We recorded the nuclear ($R(t_k)$) and bath
($Z(t_k)$) coordinates every 2~fs, and generated a total of 400
snapshots. Based on $R(t_k)$, and $Z(t_k)$, we performed a total of 16
$\times$ 400 quantum chemistry calculations, resulting in excitation
energies $\epsilon_i(t_k)$, $i = 1,...,16$, for each snapshot $k$. Based
on $R(t_k)$ and $Z(t_k)$, we also calculated $W_{i,j}(t_k)$ according to
Eq.~(\ref{eq:ddis_14}).  Combining $\epsilon_i(t_k)$ and $W_{i,j}(t_k)$
we constructed the time-dependent exciton Hamiltonian
(\ref{eq:ddis_13}).

\subsubsection{Exciton Hamiltonian}

\begin{figure}[t]
\begin{center}
\includegraphics[clip,width=3.4in]{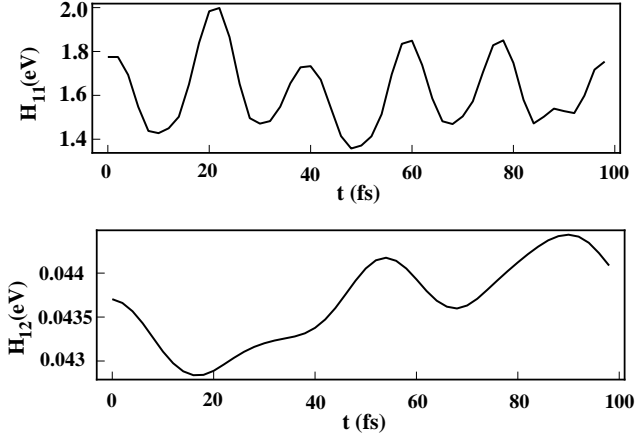}
\caption{ Matrix elements $H_{11}(t) = \epsilon_1(t)$ (top) and
$H_{12}(t) = W_{12}(t)$ (bottom) of
Hamiltonian~(\ref{eq:ddis_13}), as calculated for the first 50
snapshots of the MD run. } 
\label{fig:disH}
\end{center}
\end{figure}

Fluctuations of two representative elements of Hamiltonian matrix
(\ref{eq:ddis_13}) are presented in Fig.~\ref{fig:disH}. The
fluctuations of the largest off-diagonal matrix elements, i.e.,
couplings between neighboring BChls, are two orders of magnitude
smaller than the fluctuations of diagonal matrix elements. The
fluctuations of couplings between non-neighboring BChls are at least
another order of magnitude smaller than those of the nearest
neighbors. This suggests that off-diagonal disorder is negligible
compared to the diagonal disorder.

\begin{figure}[t]
\begin{center}
\includegraphics[clip,width=3.4in]{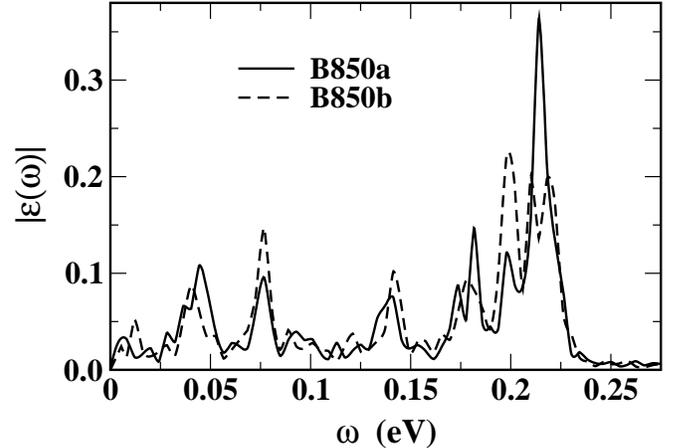}
\caption{ Average power spectrum of matrix elements $\epsilon_i(t)$, for
  B850a BChls $i = 1,3,\ldots,15$ (full line), and for B850b BChls, $i =
  2,4,\ldots,16$ (dashed line).  }  
\label{fig:vibr}
\end{center}
\end{figure}

Simple visual inspection of the time-dependence of $\epsilon(t)$ in
Fig.~\ref{fig:disH}, reveals a prominent oscillatory component of about
20~fs. Indeed, the power spectrum of $\epsilon_i(t)$ (averaged over all 16
BChls) defined as
\[
\sqrt{\frac{1}{16}\sum_{i=1}^{16} \left|\int_0^\infty dt e^{i\omega
    t}\epsilon_i(t)\right|^2}\;,
\]
and shown in Fig.~\ref{fig:vibr}, displays strong components in the
range between 0.2~eV and 0.22~eV, corresponding to the oscillations with
a period of 20.6~fs and 19.1~fs, respectively. These modes most likely
originate from a C=O or a methine bridge stretching vibration.  We note
that the autocorrelation function of Q$_y$ excitation energy of BChl in
methanol also displays a prominent vibrational mode of about 18~fs
period~\cite{MERC99}.

Fig.~\ref{fig:vibr} reveals further vibrational modes. While the B850a
and B850b BChls have some common modes (0.198~eV, 0.141~eV, 0.077~eV),
some other modes peak at slightly different energies; 0.218~eV, 0.21~eV,
0.182~eV, 0.174~eV, 0.045~eV for B850a as compared to 0.214~eV,
0.178~eV, 0.040~eV for B850b. These differences most likely stem from
the difference in the protein environment seen by B850a and B850b.

A histogram of excitation energies $\epsilon_k(t)$ of all 16
BChls, is shown in Fig.~\ref{fig:DOS}. Histograms of 8 B850a BChls
and 8 B850b BChls (data not shown) reveal rather large values for the
full-width at half-maximum (FWHM), of about 0.21~eV for B850a and
0.25~eV for B850b. Furthermore, the lineshape in
Fig.~\ref{fig:DOS} reveals a non-Gaussian distribution. The same
non-Gaussian distribution of excitation energies was observed
in~\cite{MERC99}. The reason for such a distribution might be strong
coupling to the high-frequency modes.

\begin{figure}[t]
\begin{center}
\includegraphics[clip,width=3.4in]{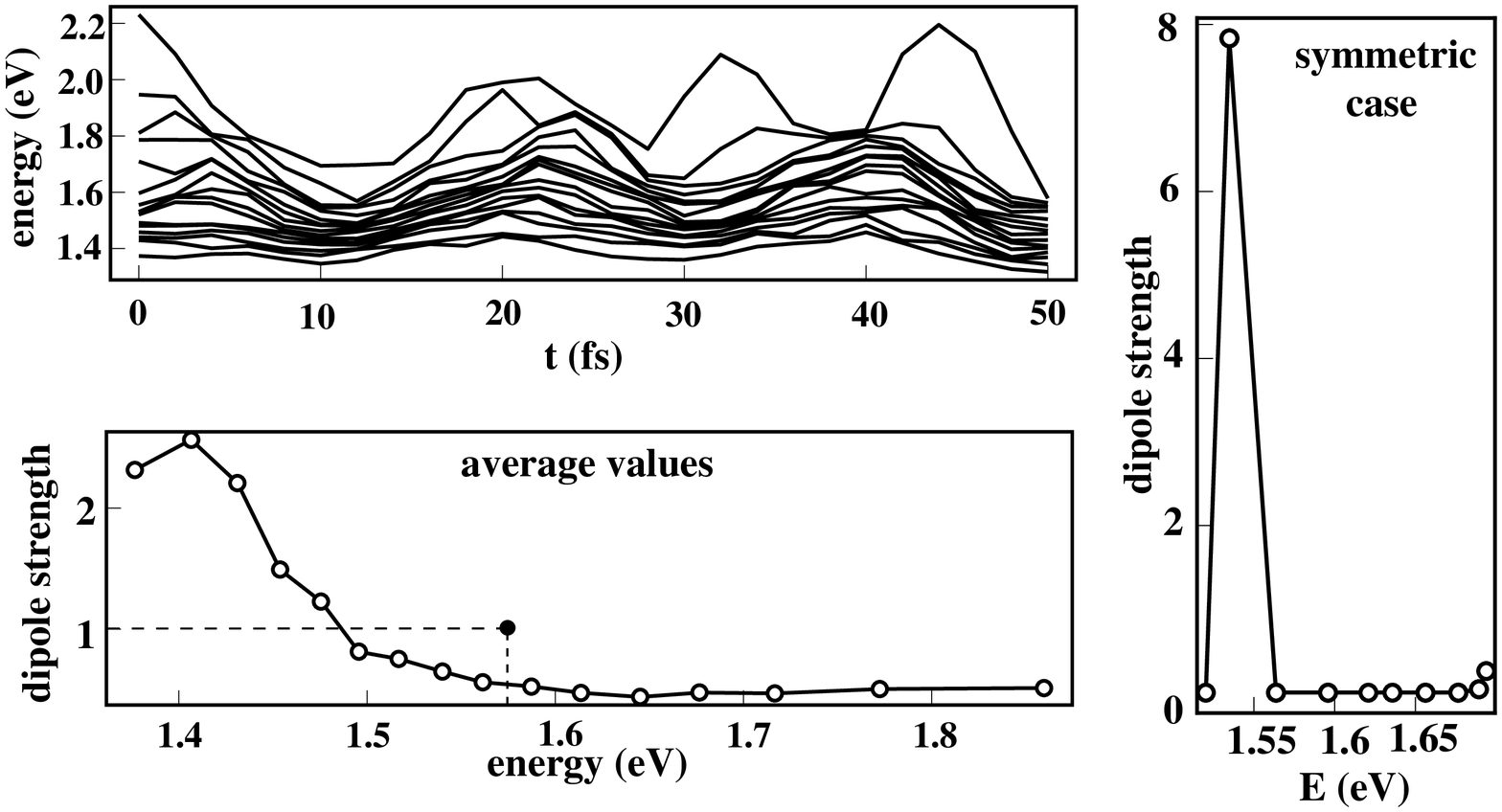}
\caption{\textbf{Top, left:} Excitonic
  energies, as determined by diagonalization of
  Hamiltonian~(\ref{eq:ddis_13}) for the first 25 snapshots of the MD
  simulation. Energies are in eV.  \textbf{Bottom, left:} Average
  energies of the 16 excitonic levels versus their corresponding average
  dipole strengths. Dipole strengths are given in units of dipole
  strength (1~$S_y$) of individual BChls. The filled circle indicates the
  excitation energy (1.57~eV) and the unit dipole strength of an
  individual BChl. \textbf{Right:} Excitonic energies for a symmetric
  (see text) Hamiltonian, versus corresponding dipole strengths.  }
\label{fig:disexc}
\end{center}
\end{figure}

The Hamiltonian~(\ref{eq:ddis_13}) was diagonalized for each snapshot
$t_k$ of the simulation, assuming that it is constant in the
interval $[t_k, \, t_{k+1}]$.  Fluctuations of the excitonic energies
that result from the diagonalization are shown in
Fig.~\ref{fig:disexc}. The Figure also shows energies of the 16
excitonic states averaged over the 800~fs long run, versus a
corresponding average of the dipole strength. Dipole strengths are
given in units of individual BChl dipole strength (1 $S_y$). As can be
inferred from Fig.~\ref{fig:disexc}, only the lowest five excitonic
levels have dipole strengths larger than 1. This suggests that in a
thermally disordered exciton system the spectral maximum experiences a
red-shift relative to the absorption maximum of an individual
BChl. 

For comparison we show excitonic energies and dipole strengths for the
Hamiltonian that corresponds to the perfect symmetric crystallographic
structure. Energetic degeneracies of the exciton states are lifted in
the fluctuating case and excitonic energies spread due to disorder. On
the other hand, dipole strength, which is mostly contained in two
degenerate excitonic levels in the symmetric case, is distributed more
equally among the exciton states in case of the disordered system. The
dipole forbidden nature of the lowest exciton state of the symmetric
case is lifted in the disordered case.

\subsubsection{Absorption spectrum}

\begin{figure}[t]
\begin{center}
    \vspace*{-8ex}
\includegraphics[clip,width=3.2in]{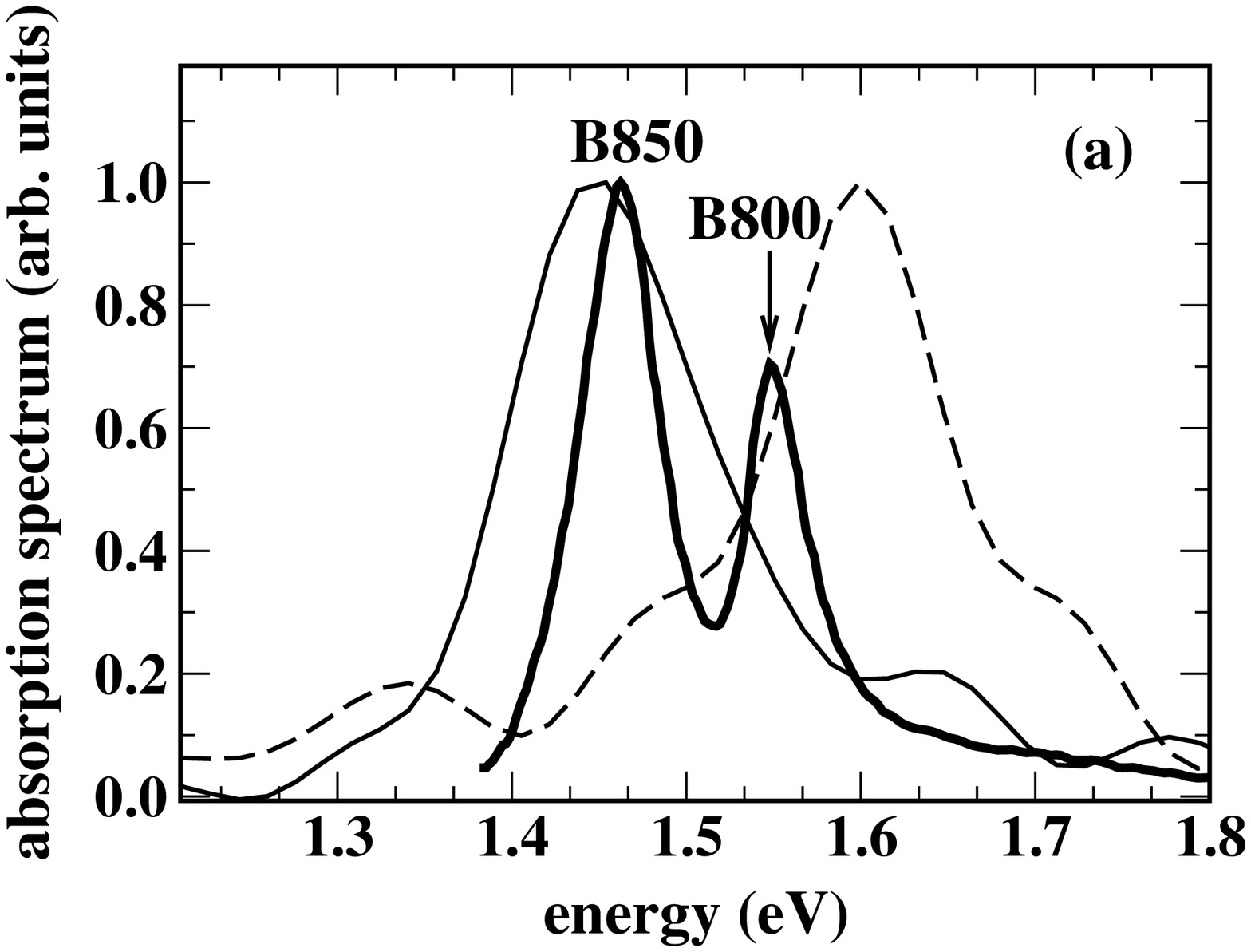}\\

\includegraphics[clip,width=3.2in]{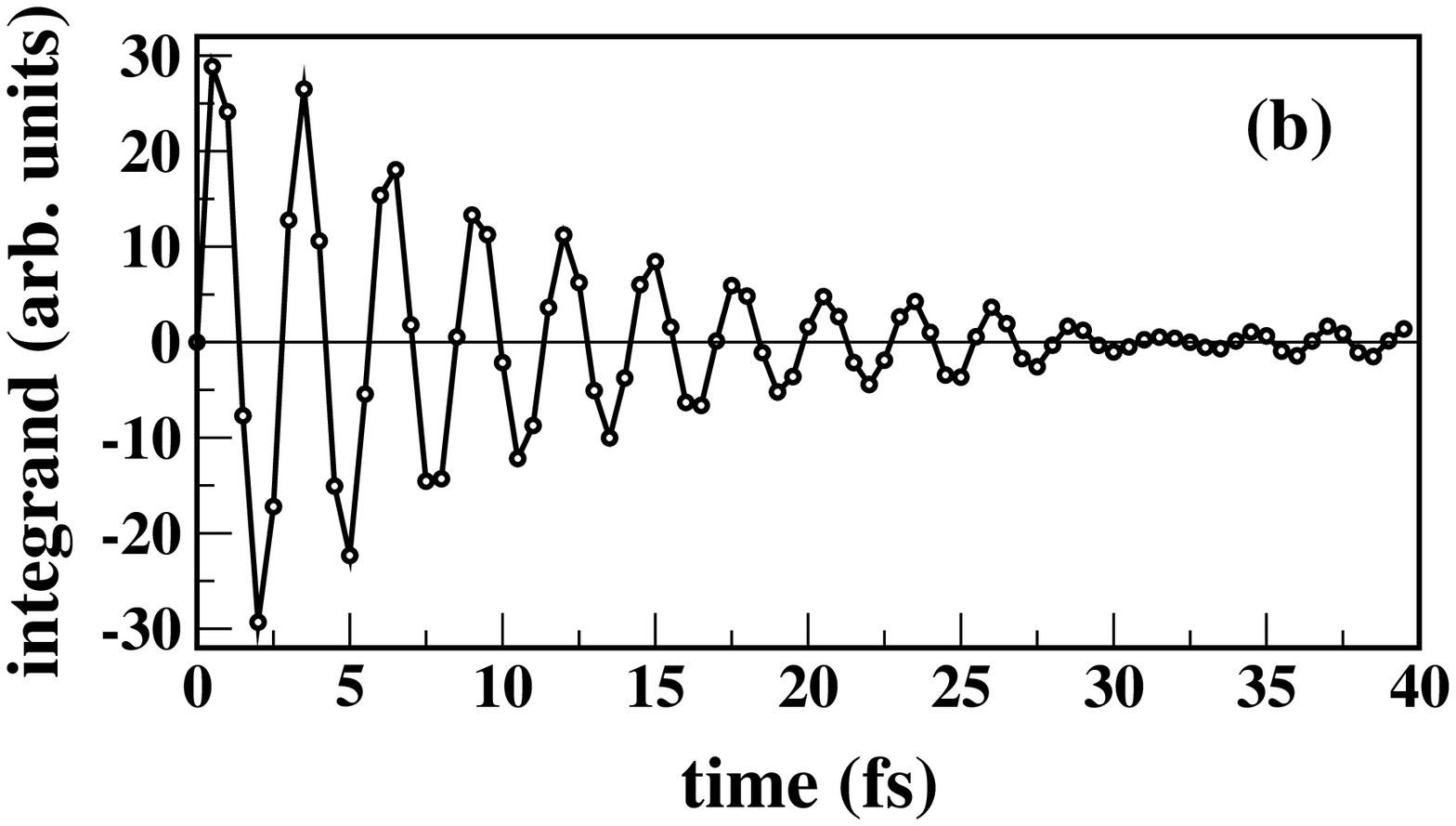}

    \vspace*{3ex}

\caption{(a) Comparison of absorption spectra of B850
  exciton (solid line) and individual BChls (dashed line), both obtained
  through Fourier transform of the time series data (see text) with the
  experimental absorption spectrum of B850 and B800 (solid thick line)
  \protect\cite{ZHAN2000}.  (b) Time-dependence of the integrand in
  Eq.~(\ref{eq:abs_19}).}
\label{fig:abs}
\end{center}
\end{figure}

The absorption coefficient $\alpha(\omega)$ of the excitonically coupled
B850 BChls was calculated by employing
Eqs.~(\ref{eq:ddis_17},\ref{eq:ddis_18},\ref{eq:abs_19}). The
Hamiltonian $\hat H(t_k)$ was assumed to be constant in the time
interval $t_{k+1}-t_k=0.5$~fs, and the evolution operators
$U(t_{k+1},t_k)$ in Eq.~(\ref{eq:ddis_17}) were determined accordingly.
Since molecular dynamics trajectories were recorded only every 2~fs we
generated additional data points for the Hamiltonian by linear
interpolation.  Ensemble averaging was achieved by dividing the total
simulation time of 800~fs into 40~fs long time intervals, which resulted
in a total of 20 samples of 40~fs length.  The integrand in
Eq.~(\ref{eq:abs_19}) is shown in Fig.~\ref{fig:abs}. The fast
fluctuation of the integrand required a choice of 0.5~fs for the time
discretization.  The resulting normalized spectrum is also shown in
Fig.~\ref{fig:abs}.

The calculated exciton absorption spectrum is compared in
Fig.~\ref{fig:abs} to the experimentally measured spectrum. The
calculated spectrum has a FWHM of about 0.138~eV, which is considerably
larger than the FWHM of 0.05~eV for the experimental spectrum. Below we
will discuss possible reasons for this discrepancy.

We have also determined the absorption spectrum of individual BChls,
according to Eq.~(\ref{eq:abs_21}). The spectrum was obtained from a
sample of 320 ($800\times 16/40$) MD runs of 40~fs length. The
integration time-step was again 0.5~fs. The FWHM of the individual BChl
absorption spectrum is about 0.125~eV, and its maximum is blue shifted
with respect to the maximum of the exciton absorption spectrum. This
shift is in agreement with the above mentioned transfer of dipole
strength into low energy exciton states (see Fig.~\ref{fig:disexc}).

Our test calculations indicate that for both the individual BChl and
exciton case, the additional peaks on both sides of the main
absorption peak are unphysical and are most likely a consequence of
insufficient sampling.

The exciton spectrum is slightly broader than the spectrum of individual
BChls, apparently indicating the lack of an exchange narrowing
of the absorption spectrum.  Exchange narrowing is well known in
J-aggregates with static disorder~\cite{SCHE96}, and is caused by
excitonic interactions within an aggregate, i.e., the excitation
experiences more than one realization of disorder, and, thus, effective
disorder is an average over those realization. For purely statically
disordered LH-II aggregates, the exchange narrowing factor was found to
be about 2.8~\cite{HU97}, in agreement with the theoretical value
$\sqrt{N}$, where $N$ characterizes the N-fold symmetry of the
aggregate, i.e., in the present case $N=8$. Exchange narrowing was also
shown to occur for dynamically disordered aggregates. For a particular
model, the authors in~\cite{WUBS98} have derived an expression for the
narrowing factor which depends on the size of the aggregate, correlation
time of the fluctuation $t_c$, as well as the strength of the excitonic
coupling $J$. For the values characteristic for our simulation, $t_c
\sim 20$~fs, and $J \sim 300$~cm$^{-1}$, the exchange narrowing factor
is about unity. Thus, provided that the model of
Ref.~\onlinecite{WUBS98} is applicable to our simulation, this result
may explain the absence of exchange narrowing of our exciton absorption
spectrum. 

\subsubsection{Discussion of the results of the simulations}

For about 30~$\%$ of the MD geometries the quantum chemistry
calculations predicted a slight degree of mixing between the Q$_y$ and
Q$_x$ states. If this mixing was only an artifact of quantum chemistry
calculations, it would have resulted in the Q$_y$ energy being predicted
too low, and fluctuation of the excitation energy to be unrealistically
large. An immediate test of the validity of the quantum chemistry
calculations is provided by a comparison of the predicted mean energy
gap of 0.45~eV between the Q$_y$ and Q$_x$ states with the corresponding
experimental value. Taking into account the additional shift due to
excitonic interactions in the Q$_y$ band, of about 0.13~eV, as predicted
by our calculations, it appears that the quantum chemistry calculations
account correctly for the experimentally measured $0.6$~eV of the mean
energy gap. More refined quantum chemistry calculations are necessary to
determine if the Q$_y\,-\,$Q$_x$ mixing is indeed genuine.

Next we discuss the choice of nearest-neighbor couplings.
Various estimates of these couplings have been reported
in the literature; for LH-II of {\em Rs.~molischianum}: 806 cm$^{-1}$
and 377 cm$^{-1}$ (semiempirical INDO/CIS calculations/effective
Hamiltonian calculations~\cite{CORY98,HU97}), 408 cm$^{-1}$ and 366
cm$^{-1}$ (collective electronic oscillator approach~\cite{TRET2000});
for LH-II of {\em Rps.~sphaeroides}: 300 cm$^{-1}$ and 233 cm$^{-1}$
(circular dichroism studies~\cite{KOLH98}); for LH-II of {\em
  Rps.~acidophila}: 238 cm$^{-1}$ and 213 cm$^{-1}$~\cite{KRUE98A}, 320
cm$^{-1}$ and 255 cm$^{-1}$~\cite{SCHO99} (CIS Gaussian 94
calculations), 394 cm$^{-1}$ and 317 cm$^{-1}$ (QCFF/PI
calculation~\cite{ALDE97}), 622 cm$^{-1}$ and 562 cm$^{-1}$ (ZINDO/S
method~\cite{LINN99}). The average values of couplings used in this
paper are 364 cm$^{-1}$ and 305 cm$^{-1}$. If larger values were used,
as might have been suggested from some of the above mentioned estimates,
a narrower absorption spectrum would have been predicted, due to an
increase in the exchange narrowing factor.

To reduce computational cost of our simulation, we employed the
dipole-dipole approximation to describe variations of the nearest
neighbor couplings as a function of time. It is well known that this
approximation does not hold in the case when center-to-center distances
between pigments are smaller than the size of the pigments themselves.
However, we find that the fluctuations of the amplitude of off-diagonal
matrix elements is negligible compared to the fluctuations of the
diagonal matrix elements, and we expect that this would still hold if
the proper coupling between the neighboring pigments was determined.

\section{Analysis in the Framework of the Polaron Model}

In order to describe the influence of dynamical disorder (thermal
fluctuations) on the electronic excitations of the B850 BChls, we will
employ now a model that focuses on the effect of the strongest
fluctuations in the exciton Hamiltonian (\ref{eq:ddis_13}), the variation
of the site energies $\epsilon_j(t)$.  These variations are foremost due
to the coupling of the high-frequency vibrations to Q$_y$ excitations in
BChl (see Fig.~\ref{fig:vibr}). Initially, we will model this coupling
through a single harmonic oscillator mode $\omega_0$, that is coupled to
each of the local BChl excitations and, thereby, to the exciton system.
Both the selected vibrational mode and the exciton system will be
described quantum mechanically, replacing the time dependence of the
Hamiltonian (\ref{eq:ddis_13}) by a time-independent Hamiltonian that
includes the selected vibrational mode and the exciton system.  We
assume that the electronic excitations of the B850 BChls form a linear
system of coupled two-level systems, which interact at each site with
dispersionless Einstein phonons of energy $\hbar \omega_0$.  To
simplify our notation, we will assume units with $\hbar\, = \, 1$.  The
corresponding model Hamiltonian, the so-called Holstein
Hamiltonian\cite{HOLS59}, reads
\begin{mathletters}
\label{eq:HH}
\begin{eqnarray}
H &=& H_{ex} + H_{ph} + H_{int}\;,\\ 
\label{eq:HHa}
H_{ex} &=& \sum_i \epsilon_i \, B_i^{\dagger}\, B_i \, + \, 
\sum_{i\neq j} V_{ij} \, B_i^{\dagger}\, B_j \,,\\
\label{eq:HHb}
H_{ph} &=& \sum_i \omega_0 \, b_i^{\dagger}\, b_i \,,\\
\label{eq:HHc}
H_{int} &=&  g \,\omega_0 \, \sum_i B_i^{\dagger}\, B_i \,
(b_i^{\dagger}\, + \, b_i) \;.
\end{eqnarray}
\end{mathletters}
Here $H_{ex}$ describes the electronic excitations of BChls;
$B_i^{\dagger}$ and $B_i$ are creation and annihilation operators
accounting for the electronic excitation of BChl$_i$ with excitation
energy $\epsilon_i$; $V_{ij}$ [c.f., Eq.~(\ref{eq:ddis_14})] encompasses
the coupling between excitations of BChl$_i$ and BChl$_j$; $H_{ph}$
represents the selected vibrations of the BChls where $b_i^{\dagger}$
and $b_i$ denote the familiar harmonic oscillator creation and
annihilation operators.  $H_{int}$ describes the interaction between
excitons and phonons, scaled by the dimensionless coupling constant $g$.

The stationary states corresponding to the Hamiltonian (\ref{eq:HH}) are
excitons ``dressed'' with a phonon cloud and are referred to as
\textit{polarons}.  Accordingly, the suggested description is called the
\textit{polaron model}.  In what follows we shall restrict ourself only
to the ``symmetric case'', i.e., we assume $\epsilon_i=\epsilon_0$ for
all sites, and we set $V_{ij}=-V\delta_{j,i\pm 1}$, where $V$ is the
nearest neighbor interaction energy.  This choice of the Hamiltonian is
motivated by the results of our simulations, which suggest that the
exciton dynamics is determined by the size- and time-scales of the
thermal fluctuations of the excitation energies, $\epsilon_i$, of
individual BChls. Indeed, as shown in Fig.~\ref{fig:disH}, the magnitude
of the fluctuations of $\epsilon_i$ is two orders of magnitude larger
than the one corresponding to the nearest neighbor coupling energies,
$V_{i,i\pm 1}$, and at least three orders of magnitude larger than the
one corresponding to the coupling energies between non-neighboring BChls
($V_{ij}$, with $j\ne i\pm 1$). Furthermore, according to
Fig.~\ref{fig:disH}, the period of oscillation of $\epsilon_i(t)$
($i=1,\ldots,16$) of a single BChl is about 20~fs, which corresponds to
a high frequency intramolecular vibronic mode of energy $\omega_0 =
1,670$~cm$^{-1}=207$~meV (see also Fig.~\ref{fig:vibr}).  In the
exciton Hamiltonian we set $\epsilon_i=\epsilon_0=1.57$~eV, and
$V=350.9$~cm$^{-1}=43.5$~meV, corresponding to an exciton bandwidth of
$4V=174$~meV.

The effect of static disorder can be incorporated into this model by
defining $\epsilon_i$ and $V_{ij}$ as random variables drawn from a
suitably chosen distribution, which best characterizes the nature of the
disorder.  In this case, in calculating different physical observables,
besides the usual thermal average, a second, configuration average needs
to be performed.

The stationary states of the Holstein polaron Hamiltonian (\ref{eq:HH})
cannot be described analytically and one needs a suitable approximation.
In general, well controlled perturbative approximations are available
only in the weak ($g\ll 1$; for a more precise criteria see below) and
strong ($g\gg 1$) coupling limits.  
Besides conventional perturbation theory, the \textit{cumulant
  expansion} method provides a convenient alternative for solving the
polaron problem. The advantage of this method is that it provides
reliable results even for arbitrary values of the coupling constant $g$
\cite{mahan}.
By using our computer simulation results, we will estimate the value of
$g$ and show that our system falls into the weak exciton-phonon coupling
regime.  Then, we will use both perturbation theory and the cumulant
expansion method to investigate within the framework of the polaron
model the effect of thermal fluctuations on the exciton bandwidth,
coherence size, and absorption line-shape.

\subsection{Evaluation of coupling constant g}

\begin{figure}[t]
\vspace*{-3ex}
\centerline{\includegraphics[clip,width=3.4in]{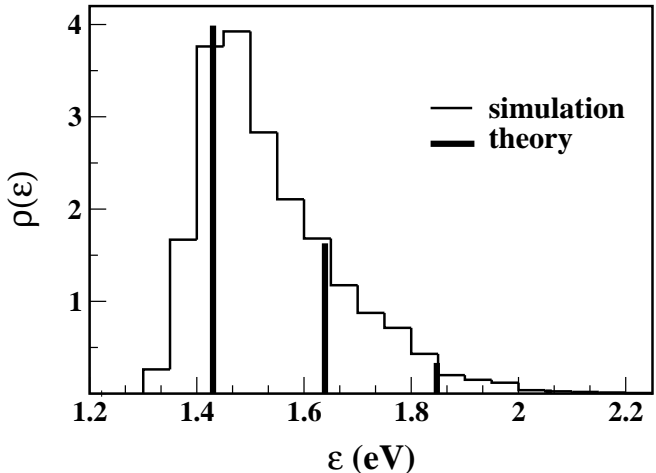}}
    \caption{Histogram of the fluctuations $\rho(\epsilon)$ of individual
      BChl excitation energies $\epsilon_i$ obtained from quantum
      chemistry calculations. The vertical bars represent the weights of the
      Dirac-delta functions in the corresponding analytical
      result~(\protect\ref{eq:DOSibm}).} 
    \label{fig:DOS}
\end{figure}

The coupling constant $g$ can be evaluated by estimating the effect of
thermal fluctuations on the electronic excitations of individual BChls.
This corresponds to setting $V_{ij}\, = \, 0$ in (\ref{eq:HH}).  In
fact, in calculating excitation energies $\epsilon_i(t)$ in our
simulations we had not accounted for excitonic coupling.  As a result,
we can use $\epsilon_i(t)$, $i=1,\ldots,M$ obtained from the
simulations, to calculate the corresponding distribution function
histogram [or density of state (DOS)] $\rho(\epsilon)=
dN(\epsilon)/d\epsilon$ (see Fig.~\ref{fig:DOS}), as well as, any of the
corresponding moments $\langle\epsilon^n_i\rangle =(1/N)
\sum_{j=1}^N\epsilon_i^n(t_j)$.  
The coupling constant $g$ can be determined by fitting to these data the
prediction from Hamiltonian (\ref{eq:HH}), for $V_{ij}\, = \, 0$.  The
statistics for $\rho(\epsilon)$ can be improved by averaging over all
sixteen B850 BChls. Since we assume that all sites are equivalent, we can
restrict ourself to the model Hamiltonian

\begin{equation}
  \label{eq:ibm}
  {\cal H} \;=\; \epsilon_0 B^{\dagger}B + \omega_0
  b^{\dagger}b + g\omega_0 B^\dagger B (b+b^{\dagger})\;,
\end{equation}
where, for brevity, we dropped the irrelevant ``i'' site index.

The DOS $\rho(\epsilon)$ can be obtained from the imaginary part of the
Fourier transform of the retarded Green's function $G_R(t) = -i
\Theta(t) \langle B(t) B^{\dagger}(0)\rangle$ \cite{mahan}, where
$\langle\ldots\rangle$ denotes the thermodynamic average over the
vibrational (phonon) degrees of freedom in the exciton vacuum.
Accordingly, we employ the identity
\begin{equation}
  \label{eq:ibm4}
  \rho(\epsilon) \;=\; -\frac{1}{\pi}{\rm Im}G_R(\epsilon)\;,
\end{equation}
where
\begin{equation}
  \label{eq:ibm5}
  G_R(\epsilon) = \int_{-\infty}^{\infty}dt\,G_R(t)e^{i\epsilon t}\;.
\end{equation}
It should be noted that, because we are dealing with a single
exciton coupled to a phonon bath, (i) the retarded Green's function will
coincide with the real time Green's function $G_R(t)=G(t)=-i\langle
TB(t)B^{\dagger}(0)\rangle$ (here $T$ denotes the time ordering
operator), and (ii) the exciton-phonon coupling will not alter the
phonon Green's function.  The latter can be written as \cite{mahan}
\begin{eqnarray}
  \label{eq:phononGF}
  D(t) = -i\langle TA(t)A(0)\rangle &=& -i [(N_0+1)\exp(-i\omega_0|t|)
  \\
  &&+ N_0 \exp(i\omega_0|t|)]\;,\nonumber
\end{eqnarray}
where
\begin{equation}
  \label{eq:ph_A}
  A(t) = b\exp(-i\omega_0 t) + b^{\dagger}\exp(i\omega_0 t)
\end{equation}
is the phonon field operator, $N_0 = 1/[\exp(\beta\omega_0)-1]$ is the
Bose-Einstein distribution function, and $\beta=1/k_BT$. At room
temperature ($T\approx 300$~K), we have $\beta\omega_0 \approx 8.1$, and
consequently $N_0\approx 3.1\times 10^{-4}\ll 1$. Thus, as we have already
mentioned, the thermal population of the high frequency vibrational mode
$\omega_0$ is negligibly small even at room temperature.

In order to calculate the Green's function we separate the Hamiltonian
(\ref{eq:ibm}) into two contributions ${\cal H} = {\cal H}_0 + {\cal
  H}_{int}$, where ${\cal H}_0$ accounts for the first term in
(\ref{eq:ibm}).  Accordingly, we express
\begin{equation}
  \label{eq:G1}
  G(t) = -i\langle T B(t)B^{\dagger}(0)\rangle = G_0(t)\exp[-\Phi(t)]\;, 
\end{equation}
where 

\begin{equation}
  \label{eq:G0}
  G_0(t) = -i\Theta(t)\exp(-i\epsilon_0 t)
\end{equation}
is the Green's function corresponding to ${\cal H}_0 $.  We then employ
the cumulant expansion method \cite{mahan} writing
\begin{equation}
  \label{eq:F_W}
  G_0(t)\exp[-\Phi(t)] \equiv
  G_0(t)\exp\left[-\sum_{m=1}^{\infty}\Phi_m(t)\right] =
  \sum_{m=0}^{\infty}W_m(t)\;, 
\end{equation}
where
\begin{eqnarray}
  \label{eq:W}
  W_m(t) &=& \frac{(-i)^{2m}}{(2m)!}\int_0^tdt_1\ldots\int_0^t dt_{2m}\\
  &&\times\langle T  B(t){\cal H}_{int}(t_1)\ldots {\cal
    H}_{int}(t_{2m})B^{\dagger}(0)\rangle\;, \nonumber
\end{eqnarray}
and ${\cal H}_{int}(t) = g\omega_0 B^{\dagger}BA(t)$. The cumulants
$\Phi_m$ are expressed in terms of $W_m$ by identifying on both sides of
Eq.~(\ref{eq:F_W}) all the terms which contain the same power of
$H_{int}$. By using Eqs.~(\ref{eq:W},\ref{eq:phononGF}) we have

\begin{eqnarray}
  \label{eq:W1}
  W_1(t) &=& -(g\omega_0)^2 G_0(t)\int_0^t dt_1 \int_0^{t_1} dt_2
  D(t_1-t_2) \\ 
  &=& G_0(t)[i\epsilon_pt -S_0 + S_+\exp(-i\omega_0t) +
  S_-\exp(i\omega_0t)]\;,\nonumber  
\end{eqnarray}
where 
\begin{equation}
  \label{eq:e_p}
  \epsilon_p=g^2\omega_0
\end{equation}
is the polaron binding energy, and

\begin{equation}
  \label{eq:SSS}
  S_0 = g^2(2N_0+1)\;,\qquad S_{\pm} = g^2(N_0+1/2\pm 1/2)\;.
\end{equation}
From these results follows
\begin{eqnarray}
  \label{eq:Phi1}
  \Phi_1(t) = - G_0^{-1}(t) W_1(t) &=& -i\epsilon_p t + S_0  - 
  S_+\exp(-i\omega_0t)\nonumber\\ 
  &&- S_- \exp(i\omega_0t) \;.
\end{eqnarray}
By employing Wick's theorem\cite{mahan}, one can calculate $W_m$, for
arbitrary integer $m>1$, with the result $W_m=G_0[G_0^{-1}W_1]^m/m!$.
Consequently, we find $\Phi_m(t)=0$ for $m>1$.  The
exact exciton Green's function is then given by

\begin{mathletters}
\label{eq:ibmGF}
\begin{eqnarray}
  \label{eq:ibmGF1}
  G(t) &=& -i\Theta(t)\exp[-i(\epsilon_0-\epsilon_p)t-\Phi_0(t)]\;,\\
  \label{eq:ibmGF2}
  \Phi_0(t) &=& S_0 - S_+\exp(-i\omega_0t) -  S_- \exp(i\omega_0t) \;.
\end{eqnarray}
\end{mathletters}
In order to calculate the DOS, we express Eq.~(\ref{eq:ibmGF2}) in the
equivalent form
\begin{eqnarray}
  \label{eq:phi0a}
  \Phi_0(t) &=&
  e^{-g^2(2N_0+1)}\exp\{2g^2\sqrt{N_0(N_0+1)}\\
  &&\times\cos[i\omega_0(t+i\beta/2)]\}\;,\nonumber
\end{eqnarray}
and employ the identity 

\begin{equation}
  \label{eq:bessel}
  e^{z\cos\theta}=\sum_{\ell=-\infty}^{\infty}{\rm
  I}_{\ell}(z)\,e^{i\ell\theta}\;, 
\end{equation}
where ${\rm I}_{\ell}(z)$ is the modified Bessel function. From
Eqs.~(\ref{eq:ibm4},\ref{eq:ibmGF}-\ref{eq:bessel}) one obtains finally
the well known result \cite{mahan,nakajima}

\begin{mathletters}
\label{eq:DOSibm}
\begin{equation}
  \label{eq:ibm8}
  \rho(\epsilon) =
  \sum_{\ell=-\infty}^{\infty}\rho_{\ell}\,\delta(\epsilon-\epsilon_{\ell})\;,
\end{equation}
with
\begin{equation}
  \label{eq:ibm9}
\rho_{\ell}=\exp[-g^2(2N_0+1)+\ell(\beta\omega_0/2)]\;
{\rm I}_{\ell}(2g^2\sqrt{N_0(N_0+1)})\;,
\end{equation}
and
\begin{equation}
  \label{eq:ibm10}
\epsilon_{\ell}=\omega_0\ell+\epsilon_0-\epsilon_p\;.
\end{equation}
\end{mathletters}
In general, when the coupling $V_{ij}$ between excitons cannot be
neglected, the higher order cumulants $\Phi_m$ do not vanish, and the
Green's function $G(t)$ cannot be calculated exactly.  Nevertheless,
even in this case, the cumulant expansion method, as illustrated above
in deriving the well known result (\ref{eq:DOSibm}), can be conveniently
applied to calculate perturbatively, in a systematic way, both the
Green's function and the corresponding absorption spectrum (see below)
to any desired degree of accuracy.

The calculated DOS (normalized to unity) is an infinite sum of
Dirac-delta functions at energies $\epsilon_{\ell}$, with weights
$\rho_{\ell}$.
As shown in Fig.~\ref{fig:DOS}, for a proper choice of the
exciton-phonon coupling constant $g$, the ``stick-DOS'' (i.e., the
weights $\rho_{\ell}$) matches well the histogram $\rho(\epsilon)$
obtained from our quantum chemistry calculations.
The coupling constant $g$ was determined by matching the first few
moments of $\epsilon$, once calculated according to
Eqs.~(\ref{eq:ibm8}-\ref{eq:ibm10}), and then from the time series
$\{\epsilon_i(t)\}_{i=1,...,16}$. The definition of the moments
\begin{equation}
  \label{eq:ibm11}
  \langle\epsilon^{n}\rangle =
  \int_{-\infty}^{\infty}d\epsilon\rho(\epsilon)\epsilon^n =
  \sum_{\ell}\rho_{\ell} \epsilon_{\ell}^n\;,   
\end{equation}
applied to (\ref{eq:ibm8}-\ref{eq:ibm10}), results in analytical
expressions that can be compared with the numerical values
determined from the time series.  One obtains
\begin{mathletters}
\begin{eqnarray}
  \label{eq:ibm12a}
  \langle\epsilon\rangle &=& \epsilon_0 \approx 1.57~{\rm eV}\;, \\
  \label{eq:ibm12b}
  \langle\epsilon^2\rangle &=&
  \langle\epsilon\rangle^2+g^2(2N_0+1)\omega_0\approx
  2.32~{\rm eV}^2\;, \\ 
  \label{eq:ibm12c}
  \langle\epsilon^3\rangle &=& \langle\epsilon\rangle^3 +g^2\omega_0^3
  \left[1+3g^2(2N_0+1)\right.\\
  &&\left.+3(2N_0+1)\left(\frac{\epsilon_0}{\omega_0}-g^2\right)\right]
  \approx 3.56~{\rm eV}^3 \;.\nonumber
\end{eqnarray}
\end{mathletters}
The coupling constant can be obtained from Eq.~(\ref{eq:ibm12b})

\begin{equation}
  \label{eq:ibm13}
  g = \sqrt{\frac{\langle\epsilon^2\rangle -
  \langle\epsilon\rangle^2}{(2N_0+1)\omega_0^2}} \approx 0.65\;,
\end{equation}
Equation~(\ref{eq:ibm12c}) now contains no adjustable parameters and it can
be used to check the reliability of our approach. It is found that the
difference between the two sides of this equation is less than $2\%$.

The value of $g$, given by (\ref{eq:ibm13}), does not determine by
itself the value of the ratio between $H_{int}$ ($\sim\epsilon_p$) and the
problematic hopping term ($\sim 4V$, i.e., the energy bandwidth of the
excitons) in the Holstein Hamiltonian. The actual dimensionless coupling
strength parameter for the polaron model is $\kappa\equiv\epsilon_p/4V
=g^2\omega_0/4V \approx 0.5$.  This value corresponds to a weak coupling
regime of the polaron model \cite{mahan,nakajima}.

\subsection{Polaron bandwidth}

We return now to the full Holstein Hamiltonian (\ref{eq:HH}) with
$V_{ij}\, \ne \, 0$. In the weak coupling limit it is convenient to
rewrite the polaron Hamiltonian (\ref{eq:HH}) in ``momentum space'' as

\begin{mathletters}
\label{eq:HH_k}
\begin{eqnarray}
  \label{eq:HH_kH}
  H &=& H_0 + H_{int}\;,\\
  \label{eq:HH_k0}
  H_0 &=& \sum_k \epsilon_kB^{\dagger}_kB_k + \omega_0\sum_q
  b^{\dagger}_qb_q\;,\\
  \label{eq:HH_ki}
  H_{int} &=& \frac{g\omega_0}{\sqrt{M}}\sum_{k,q} B^{\dagger}_{k+q} B_k
  (b^{\dagger}_q+b_{-q})\;,
\end{eqnarray}
\end{mathletters}
where $B_k=(1/\sqrt{M})\sum_j B_j\exp(ikj)$, etc.  For
$V_{ij}=-V\delta_{j,i\pm 1}$ ($V>0$), the exciton dispersion (measured
from the site excitation energy $\epsilon_0$) is
$\epsilon_k=-2V\cos(k)$, with $k\equiv k_m=2\pi m/M$,
$m=0,1,\ldots,M-1$, and $q\equiv q_p=2\pi p/M$, $p=0,1,\ldots,M-1$.

The polaron (phonon renormalized exciton) spectrum $E_k$ can be
calculated by using second order perturbation theory (all odd order
terms vanish because the phonon creation and annihilation operators must
appear in pairs)

\begin{equation}
  \label{eq:pol_spec}
  E_k = \epsilon_k +
  \frac{g^2\omega_0^2}{M}\sum_q\frac{1}{\epsilon_k-\epsilon_{k+q}-\omega_0}\;.
\end{equation}
The renormalized exciton wave functions to leading order in $g$
are given by

\begin{equation}
  \label{eq:pol_wf}
  |k;0) = |k;0\rangle +\frac{g\omega_0}{\sqrt{M}}\sum_q
   \frac{1}{\epsilon_k - \epsilon_{k+q}-\omega_0}|k+q;1_q\rangle\;.
\end{equation}
Here $|k;0\rangle=(1/\sqrt{M})\sum_m\exp(ikm)|m\rangle$ denotes the
exciton wave function (in site $|m\rangle$ representation) in the
absence of the phonons, while $|k+q;1_q\rangle$ represents a state of
non-interacting exciton (with momentum $k+q$) and phonon (with momentum
$q$). The applicability of perturbation theory is subject to the
condition $g\omega_0/\sqrt{M}\ll |\epsilon_k -
\epsilon_{k+q}-\omega_0|$, which in our case is marginally fulfilled
since $g\omega_0/\sqrt{M} = 33$~meV$<\text{min}|\epsilon_k
-\epsilon_{k+q}-\omega_0|=35$~meV.

According to Eq.~(\ref{eq:pol_spec}) the renormalized exciton (or
polaron) bandwidth is 

\begin{equation}
  \label{eq:dEp}
  \Delta E_p \equiv \text{max}[E_k]-\text{min}[E_k] \approx
  66\text{meV}\;,
\end{equation}
which represents a $60$\% reduction with respect to the unperturbed
bandwidth $4V$. Thus the effect of the high frequency phonons is to
reduce the exciton energy band, a situation commonly encountered in
other weak coupling polaron models in which the phonons tend to enhance
the quasiparticle mass, which is translated into a reduction of the
width of the corresponding energy band \cite{mahan,nakajima}.  However,
it is known that static disorder (or coupling to low frequency phonons)
has precisely the opposite effect of increasing the exciton bandwidth in
BChl aggregates. The fact that in our MD simulations we observe an
increase instead of a decrease of the renormalized exciton bandwidth is
most likely due to the fact that in our simplified polaron model the
effect of static disorder is entirely neglected, while in the MD
simulations this is implicitly included.

\subsection{Polaron coherence length}

Next, we determine the effect of the phonons on the coherence length
$L_{\rho}$ of the excitons in the BChl ring, by using our polaron model.
At zero temperature, the excitons in the coupled BChl aggregate are
completely delocalized, and $L_{\rho}$ should coincide with the system
size $M$. In general, finite temperature and any kind of disorder will
reduce the coherence length of the exciton.  There is no universally
accepted definition of $L_{\rho}$. We employ here an expression based on
the concept of inverse participation ratio, familiar from quantum
localization~\cite{economou,MEIE97}

\begin{equation}
  \label{eq:L_rho}
  L_{\rho} = \left[\left(\sum_{ij}|\rho_{ij}|\right)^2\right]
  \left[M\sum_{ij}|\rho_{ij}|^2\right]^{-1} \;,
\end{equation}
where the reduced exciton density matrix is given by

\begin{equation}
  \label{eq:d_mat}
  \rho_{ij} = \sum_k C_k^*(i)C_k(j) \exp(-\beta E_k)\;.
\end{equation}
In the absence of the exciton-phonon coupling, at room temperature, by
setting in the above equations $C_k(j)=\exp(ikj)/\sqrt{M}$ and
$E_k\equiv\epsilon_k = -2V\cos(k)$, one obtains $L_{\rho 0}\approx 6.4$,
which represents a dramatic decrease with respect to the corresponding
zero temperature value of 16.
For weak exciton-phonon coupling, the effect of the phonons on
$L_{\rho}$ can be taken into account via perturbation theory. By
employing Eqs.~(\ref{eq:pol_spec},\ref{eq:pol_wf}) in the density matrix
(\ref{eq:d_mat}), Eq.~(\ref{eq:L_rho}) yields $L_{\rho}\approx 5.4$. As
expected, the phonons, which act as scatterers for the excitons, reduce
the coherence size of the latter.

\begin{figure}[t]
  \begin{center}
    \hspace*{2ex}\includegraphics[clip,width=3.2in]{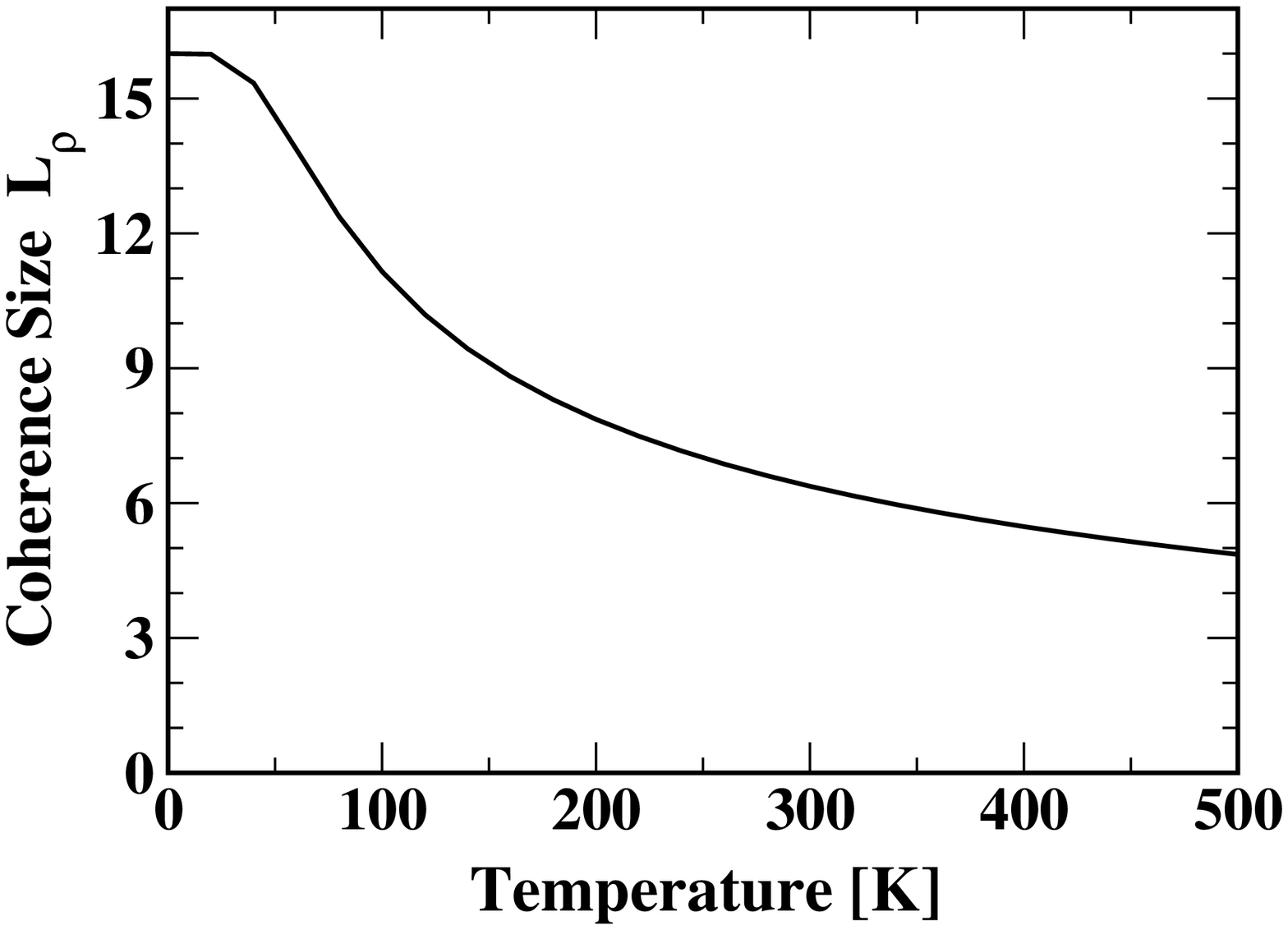}

    \vspace*{3ex}

    \includegraphics[clip,width=3.1in]{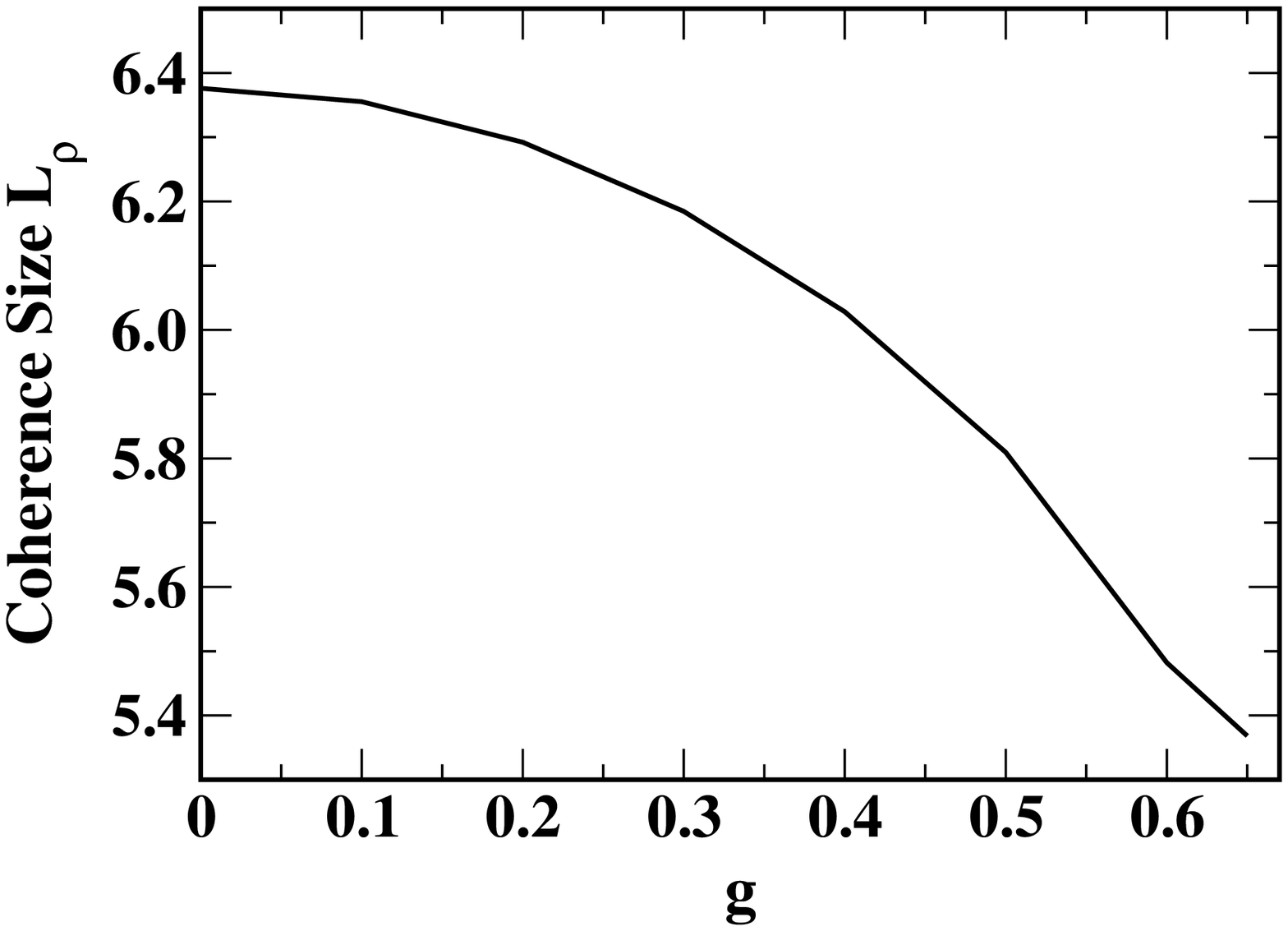}

    \vspace*{3ex}

    \caption{\textbf{Top}: Temperature dependence of the exciton
      coherence length in the absence of disorder. \textbf{Bottom}:
      Polaron coherence length as a function of the exciton-phonon
      coupling constant $g$ at room temperature.}
    \label{fig:Lr-Tg}
  \end{center}
\end{figure}

As it can be inferred from Fig.~\ref{fig:Lr-Tg}, the localization of the
exciton is due primarily to thermal averaging and to a lesser extent to
dynamic disorder. This conclusion is in agreement with previous studies
\cite{RAY99}.

\subsection{Absorption spectrum for single Einstein phonon}
\label{sec:pol_abs1}

Finally, we discuss the effect of the phonons on the absorption spectrum
of the excitons.  The fundamental absorption spectrum $I(\omega)$
(line shape function) is defined as [cf.~Eq.~(\ref{eq:abs_21})]

\begin{equation}
  \label{eq:I3}
  I(\omega) = \frac{1}{2\pi}\int_{-\infty}^{\infty}dt f(t)e^{i\omega t}\;,
\end{equation}
where the generating function $f(t)$, up to irrelevant factors, is given
by \cite{nakajima}

\begin{equation}
  \label{eq:I1}
  f(t) = \sum_k |d_k|^2 e^{-i\epsilon_k t}\langle\langle
  k|U(t)|k\rangle\rangle\;. 
\end{equation}
Here $|d_k|$ is the magnitude of the transition dipole moment
connecting the ground electronic state and the $|k\rangle$ exciton
state; $\langle\langle k|\ldots|k\rangle\rangle$ denotes the thermal
average over the phonons of the excitonic matrix element $\langle
k|\ldots|k\rangle$, and $U(t)$ is an evolution operator given by

\begin{eqnarray}
  \label{eq:I2}
  U(t) &=& \exp(iH_0t)\exp[-i(H_0+H_{int})t]\\
  &=& T\exp\left[-i\int_0^t
  d\tau H_{int}(\tau)\right]\;,\nonumber
\end{eqnarray}
with 
\begin{mathletters}
\begin{eqnarray}
  \label{eq:I2a}
  H_{int}(t)&=&\exp(iH_0t)H_{int}\exp(-iH_0t)\\
  &=& g\omega_0 \sum_{q} \rho_q(t)A_q(t) \;,\nonumber
\end{eqnarray}
expressed in the interaction representation. The time dependent exciton
density operator is given by
\begin{equation}
  \label{eq:I2b}
\rho_q(t)=\sum_k e^{-i(\epsilon_k-\epsilon_{k+q})t} B_{k+q}^{\dagger}B_k\;,   
\end{equation}
while the phonon field operator reads
\begin{equation}
  \label{eq:I2c}
  A_q(t) = b_q e^{-i\omega_0t} + b_{-q}^{\dagger} e^{i\omega_0t}\;.
\end{equation}
\end{mathletters}
The generating function (\ref{eq:I1}) can be calculated by using the
cumulant expansion method discussed above. Indeed, we can write in
analogy to Eqs.~(\ref{eq:F_W},\ref{eq:W})

\begin{eqnarray}
  \label{eq:IU_t}
  \langle\langle k|U(t)|k\rangle\rangle \equiv
  \exp[-\Phi_{k}(t)] &=& 
  \exp\left[-\sum_{m=1}^{\infty}\Phi_m(t)\right]\\
  &=& \sum_{m=0}^{\infty}W_m(t)\;,\nonumber
\end{eqnarray}
where 

\begin{eqnarray}
  \label{eq:IWm}
  W_m(t) = \frac{(-i)^{2m}}{(2m)!}&&\int_0^t dt_1 \ldots \int_0^t
  dt_{2m}\\
  &&\times\langle\langle k| T H_{int}(t_1)\ldots
  H_{int}(t_{2m})|k\rangle\rangle\;. \nonumber
\end{eqnarray}
As for the Green's function (\ref{eq:F_W}), $W_m(t)$ contains only even
numbers of time ordered $H_{int}(t)\propto A(t)$ operators, because the
average over odd numbers of phonon field operators $A(t)$ vanishes.
Similarly to Eq.~(\ref{eq:W1}), we have

\begin{eqnarray}
  \label{eq:IW1}
  W_1(t) &=& -\frac{(g\omega_0)^2}{2M} \int_0^t dt_1 \int_0^t dt_2
  \sum_{q_1,q_2} \langle TA_{q_1}(t_1)A_{q_2}(t_2)\rangle\nonumber\\
  &&\times \langle k|
  \rho_{q_1}(t_1) \rho_{q_2}(t_2)| k\rangle\nonumber\\
  &=& -\frac{(g\omega_0)^2}{2M} \int_0^t dt_1 \int_0^t dt_2
  \sum_q iD(t_1-t_2) \nonumber\\
  &&\times\langle k|\rho_{q}(t_1)
  \rho_{-q}(t_2)| k\rangle\;. 
\end{eqnarray}
A straightforward calculation yields
\begin{eqnarray}
  \label{eq:FF}
  F_k(t_1-t_2)&\equiv&\frac{1}{M}\sum_q\langle k|\rho_{q}(t_1)
  \rho_{-q}(t_2)| k\rangle \\
  &=& \frac{1}{M}\sum_q
  \exp[i(\epsilon_ k-\epsilon_{ k-q})(t_1-t_2)] \;.  \nonumber
\end{eqnarray}
Inserting Eq.~(\ref{eq:FF}) into (\ref{eq:IW1}), and using the phonon
Green's function (\ref{eq:phononGF}), the cumulant $\Phi_1(t)$ can be
written 

\begin{equation}
  \label{eq:IPhi1}
  \Phi_1(t)=-W_1(t)= g^2\omega_0^2\int_0^t d\tau (t-\tau) iD(\tau) F_k(\tau)\;.
\end{equation}
If one neglects the coupling between the individual excitations (case
corresponding to light absorption by individual BChls), by setting
$V=0$, Eq.~(\ref{eq:IPhi1}) yields the same expression as
(\ref{eq:Phi1}). Note that in this case, since $\epsilon_k=\epsilon_0$
is $k$ independent, we have $F(\tau)=1$, and similarly to the
calculation of the Green's function (\ref{eq:ibmGF}), it can be shown
that $W_m(t)=[W_1(t)]^m/m!$, for $m>2$. As a result, all the
corresponding higher order cumulants $\Phi_m$ vanish, i.e.,
$\Phi(t)=\Phi_1(t)$, and the exact expression of the line-shape function
$I_0(\omega)$ for individual BChls assumes the familiar form
\cite{mahan,nakajima}
\begin{mathletters}
  \label{eq:II0}
\begin{eqnarray}
  \label{eq:II01}
  I_0(\omega) &\propto& \exp(-S_0)\int_{-\infty}^{\infty}dt
  e^{i(\omega-\epsilon_0+\epsilon_p)t}\\
  &&\times\exp(-S_+e^{-i\omega_0t} -
  S_-e^{i\omega_0t})\nonumber\\
  \label{eq:II02}
  &=& \sum_{\ell=-\infty}^{\infty}\rho_{\ell}\delta(\omega-\omega_{\ell})\;,
\end{eqnarray}
\end{mathletters}
where $\rho_{\ell}$ and $\omega_{\ell}$ are given by
Eqs.~(\ref{eq:ibm9}), and (\ref{eq:ibm10}), respectively. In other
words, the absorption line-shape function of an exciton in the phonon
field is given by the imaginary part of the exciton Green's function,
which is proportional to the density of states \cite{nakajima}.

In general, $I(\omega)$ cannot be calculated exactly. However, even for
arbitrary values of the exciton-phonon coupling constant $g$, the cumulant
approximation $\Phi_k(t)\approx\Phi_1(t)$ can be used safely to evaluate
the generating function $f(t)$ and the corresponding line shape function
$I(\omega)$. We have

\begin{equation}
  \label{eq:abs1}
  \Phi_k(t)\approx \int_0^t d\tau (t-\tau){\cal D}(\tau) F_k(\tau)\;,
\end{equation}
where, according to Eq.~(\ref{eq:phononGF}), 
\begin{eqnarray}
  \label{eq:abs2}
  {\cal D}(t) &\equiv& g^2 \omega_0^2 iD(t) \\
  &=& g^2 \omega_0^2
  [(N_0+1)e^{-i\omega_0 t} + N_0 e^{i\omega_0 t}]\;.\nonumber
\end{eqnarray}
The generating function is

\begin{equation}
  \label{eq:abs3}
  f(t) = \sum_k |d_k|^2 e^{-i\epsilon_k t} e^{-\Phi_k(t)}\;,
\end{equation}
and the corresponding line shape function is given by Eq.~(\ref{eq:I3}).
Clearly, the coupling of the 16 exciton levels to a single Einstein
phonon $\omega_0$ leads to a \textit{stick} absorption spectrum, i.e., a
series of Dirac delta functions with different weights.

\subsection{Absorption spectrum for distribution of phonons}
\label{sec:pol_abs2}

In order to calculate, in the framework of the polaron model, the
broadening of the absorption spectrum, and to compare it with the
corresponding experimental result [see Fig.~\ref{fig:abs}], one needs to
include the coupling of the excitons to the quasi-continuous
distribution of the rest of the phonons. Formally, this can be achieved
by replacing ${\cal D}(t)$ in Eq.~(\ref{eq:abs1}) with [cf.
Eq.~(\ref{eq:abs2})]

\begin{equation}
  \label{eq:abs2a}
  {\cal D}(t) = \sum_\alpha g^2_\alpha \omega^2_\alpha iD_\alpha(t) =
  \int_0^\infty d\omega J(\omega)\, iD_\omega(t)\;,
\end{equation}
where
\begin{equation}
  \label{eq:iDw}
  iD_\omega(t) = (N_\omega +1)\exp(-i\omega t) + N_\omega\exp(i\omega t)\;.
\end{equation}
Here we have introduced the \textit{phonon spectral function}

\begin{equation}
  \label{eq:abs4}
  J(\omega) = \sum_\alpha g^2_\alpha \omega^2_\alpha
  \delta(\omega-\omega_\alpha)\;. 
\end{equation}
Once the actual form of $J(\omega)$ is known, $I(\omega)$ can be
calculated using
Eqs.~(\ref{eq:I3},\ref{eq:abs3},\ref{eq:abs1},\ref{eq:abs2a}).

Apparently, the determination of $J(\omega)$ requires the seemingly
unattainable knowledge of the energies $\omega_\alpha$ of \textit{all}
phonons, together with their corresponding coupling constants
$g_\alpha$.  However, the same problem posed itself in the framework of
the spin-boson model description of the coupling between protein motion
and electron transfer processes \cite{XU94}, and could be solved then
through a spectral function evaluated from the energy gap fluctuations
$\delta{\epsilon}(t)$. Likewise, here we can determine $J(\omega)$ from
the fluctuations of the excitation energies $\delta{\epsilon}(t)$
calculated for individual BChls.  The latter quantity can be obtained
from the combined MD/quantum chemistry simulation carried out in this
study.

\begin{figure}[t]
  \begin{center}
    \includegraphics[clip,width=3.2in]{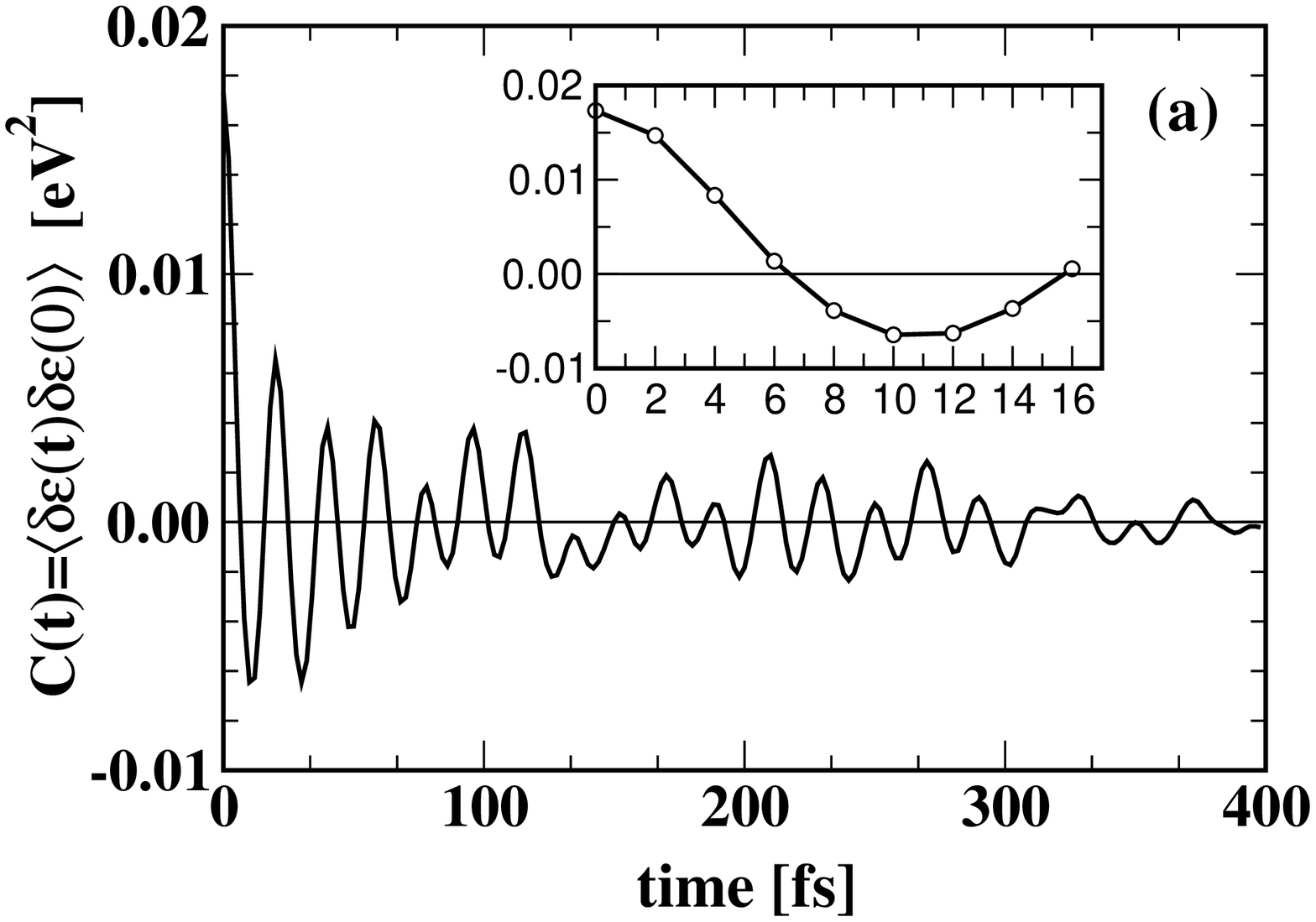}

    \vspace*{3ex}

    \includegraphics[clip,width=3.1in]{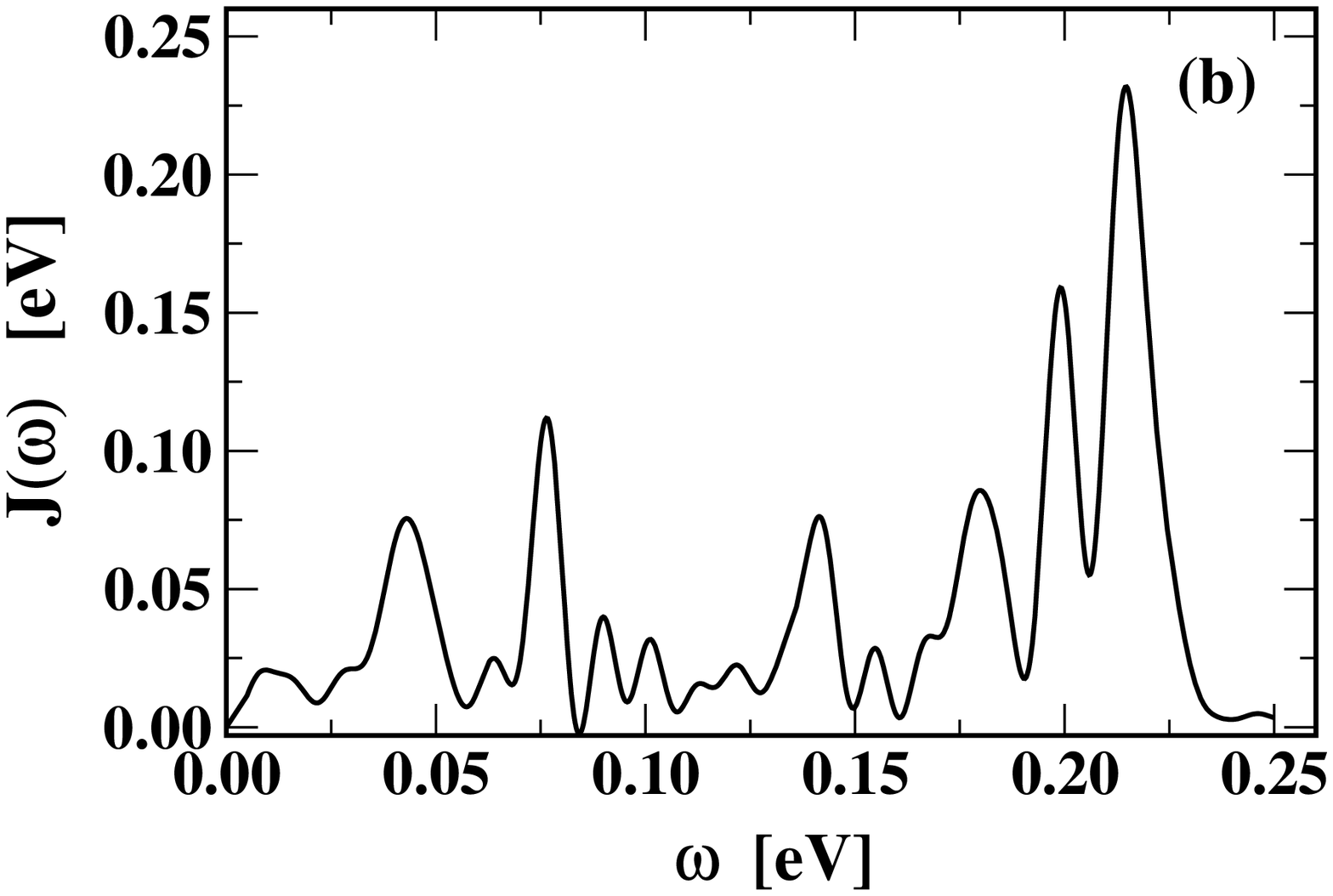}

    \vspace*{3ex}

    \caption{\textbf{(a)} Autocorrelation function ${\cal C}(t)$ of the
      energy gap fluctuations $\delta\epsilon(t)$ for individual BChls,
      calculated using Eq.~(\protect\ref{eq:C_t}). The inset shows
      the short time behavior of ${\cal C}(t)$. \textbf{(b)} Phonon
      spectral function $J(\omega)$ obtained according to
      Eq.~(\protect\ref{eq:w2J}).}
    \label{fig:corr-J}
  \end{center}
\end{figure}

Indeed, the Hamiltonian for an individual BChl interacting with a
phonon bath can be written

\begin{equation}
  \label{eq:ex-ph}
  H=H_{0}+H_{int} = (\epsilon_0 + \delta\hat{\epsilon})B^\dagger B\;,
\end{equation}
where we defined the phonon induced energy-gap-fluctuation operator 

\begin{equation}
  \label{eq:de}
  \delta\hat{\epsilon}(t) = \sum_\alpha g_\alpha\omega_\alpha A_\alpha(t)\;.
\end{equation}
The autocorrelation function of the energy gap $\delta\epsilon(t)$ is
the real part of the autocorrelation function of the
energy-gap-fluctuation operator $\delta\hat{\epsilon}(t)$, i.e.,

\begin{eqnarray}
  \label{eq:e-corr}
  {\cal C}(t)&\equiv&
  \langle\delta\epsilon(t)\delta\epsilon(0)\rangle = \text{Re}
  [\langle\delta\hat\epsilon(t)\delta\hat\epsilon(0)\rangle] \\
  &=& \text{Re}\left[\sum_\alpha g^2_\alpha
  \omega^2_\alpha\,iD_\alpha(t)\right]  
  =  \text{Re}[{\cal D}(t)]\;,\nonumber
\end{eqnarray}
where we have used Eqs.~(\ref{eq:de}) and (\ref{eq:abs2a}). Inserting
Eq.~(\ref{eq:iDw}) into (\ref{eq:e-corr}), one obtains

\begin{equation}
  \label{eq:e-corr2}
  {\cal C}(t) = \int_0^\infty d\omega J(\omega) 
  \coth(\beta\omega/2)\cos\omega t\;.
\end{equation}
$J(\omega)$ can be obtained through the inverse cosine transform of
${\cal C}(t)$, i.e.,

\begin{equation}
  \label{eq:w2J}
  J(\omega) = \frac{2}{\pi}\tanh(\beta\omega/2)\int_0^\infty\!dt
  \,{\cal C}(t)\,\cos\omega t\;.
\end{equation}
The autocorrelation function ${\cal C}(t)$ can be evaluated numerically
from the finite time series $\delta\epsilon_j(t_i)$. Here
$j=1,\ldots,16$ is the index labeling individual BChls, and
$t_i=(i-1)\times 2$~fs, $i=1,\ldots,400$, denotes the time at which the
energy gap was determined. As mentioned in the previous sections, each
time series consisted of $N=400$ time steps of 2~fs. For best sampling,
one averages over all $M=16$ BChls resulting in the time series

\begin{equation}
  \label{eq:C_t}
  {\cal C}(t_i) = \frac{1}{M}\sum_{j=1}^{M} \left[ \frac{1}{N-i}
  \sum_{k=1}^{N-i} \delta\epsilon_j(t_i+t_k)\delta\epsilon_j(t_k) \right]\;,
\end{equation}
which is shown in Fig.~\ref{fig:corr-J}a. It is safe to assume that
Eq.~(\ref{eq:C_t}) is reliable for $t\lesssim 400$~fs. We note that
the autocorrelation function in Fig.~\ref{fig:corr-J}a, resembles the
transition frequency autocorrelation function for Nile
Blue~\cite{OHTA2001}, obtained through a model that combines in both
the intramolecular vibrations and solvent dependent contributions.

The corresponding phonon spectral function $J(\omega)$ can be computed
by employing Eq.~(\ref{eq:w2J}). The result is shown in
Fig.~\ref{fig:corr-J}b. The shape of $J(\omega)$ resembles the power
spectrum of the BChl excitation energies (see Fig.~\ref{fig:vibr}). In
particular, the prominent peak around $\omega_0 \approx 0.2$~eV in
$J(\omega)$ indicates a strong coupling of the system to an
intramolecular C=O vibronic mode. Note, however, that $J(\omega)$ has
significant contributions over the entire range of phonon energies
$0<\omega \lesssim 0.22$~eV and, therefore, we expect that all, i.e.,
low, intermediate and high frequency phonons will contribute to the
broadening of the line shape function $I(\omega)$.

By assuming that the coupling of BChl molecules to phonons
(intramolecular vibrations, vibrations of the protein matrix and
solvent molecules) is independent of their excitonic coupling, we may
conclude that $J(\omega)$ given by Eq.~(\ref{eq:w2J}), describes
equally well both individual BChls and excitonic aggregates of
BChls. The only difference between the corresponding absorption
spectra comes from the ``excitonic'' factor $F_k(t)$ in
Eq.~(\ref{eq:abs1}).

In order to calculate the line shape function $I(\omega)$, as given by
Eqs.~(\ref{eq:I3},\ref{eq:abs1},\ref{eq:abs3}), from the 
energy gap autocorrelation function ${\cal C}(t)$ one proceeds as
follows. First, we determine 

\begin{eqnarray}
  \label{eq:Dt}
  {\cal D}(t) &\equiv& {\cal D}_1(t)-i{\cal D}_2(t)\\
  &=& \int_0^\infty
  d\omega J(\omega) [\coth (\beta\omega/2) \cos\omega t -i \sin\omega
  t]\;,\nonumber 
\end{eqnarray}
from which, by taking into account Eqs.~(\ref{eq:e-corr2}) and (\ref{eq:w2J}),
one obtains

\begin{mathletters}
  \label{eq:D12}
  \begin{equation}
    \label{eq:D1}
    {\cal D}_1(t) =  {\cal C}(t)\;,
  \end{equation}
and 
\begin{equation}
  \label{eq:D2}
  {\cal D}_2(t) = \int_0^\infty d\omega J(\omega) \sin\omega t\;. 
\end{equation}
\end{mathletters}
Next, we calculate the cumulant $\Phi_k(t)$; according to Eq.~(\ref{eq:abs1})

\begin{mathletters}
\label{eq:Phi12+}
\begin{equation}
  \label{eq:Phi12}
  \Phi_k(t) = {\Phi'}_k(t) - i {\Phi''}_k(t)\;,
\end{equation}
with
\begin{equation}
  \label{eq:Phi1a}
  {\Phi'}_k(t) = \int_0^\infty d\tau (t-\tau) [{\cal
  C}(\tau)\text{Re}\{F_k(t)\} - {\cal D}_2(\tau) \text{Im}\{F_k(t)\}]\;,
\end{equation}
and 
\begin{equation}
  \label{eq:Phi2a}
  {\Phi''}_k(t) = \int_0^\infty d\tau (t-\tau) [{\cal D}_2(\tau) 
  \text{Re}\{F_k(\tau)\} + {\cal C}(\tau)\text{Im}\{F_k(\tau)\}]\;.
\end{equation}
\end{mathletters}
Finally, the line shape function is according to
Eqs.~(\ref{eq:I3},\ref{eq:abs3}) 

\begin{mathletters}
\begin{equation}
  \label{eq:Iwk}
  I(\omega) \propto \sum_k |d_k|^2 \int_0^\infty dt \exp[-{\Phi'}_k(t)]
  \cos[(\omega -\epsilon_k)t + {\Phi''}_k(t)]\;.
\end{equation}

This result applies to the general case when the system has several
optically active levels, characterized by different transition dipole
moments $d_k$. In both cases considered by us, i.e., individual BChls
and the fully symmetric B850 system (i.e., a circular aggregate of 16
BChls with excitonic coupling, having 8-fold symmetry, and transition
dipole moments oriented in the plane of the ring of BChls), there is
in fact only one optically active level and, therefore, in
Eq.~(\ref{eq:Iwk}) the summation over $k$, as well as the transition
dipole moment can be both dropped. For an individual BChl we take
$\epsilon_k\equiv\epsilon_0 \approx 1.6$~eV, while for the B850 system
the optically active doubly degenerate level is $\epsilon_k\equiv
\epsilon_{\pm 1}=\epsilon_0 - 2V \cos(\pi/8) = 1.52$~eV. In this case,
the line shape function assumes the simpler form

\begin{equation}
  \label{eq:Iw_}
  I(\omega) \propto \int_0^\infty dt \exp[-{\Phi'}_k(t)]
  \cos[(\omega -\epsilon_k)t + {\Phi''}_k(t)]\;.
\end{equation}
\end{mathletters}

\begin{figure}[t]
  \begin{center}

    \vspace*{-9ex}
      \includegraphics[clip,width=2.8in]{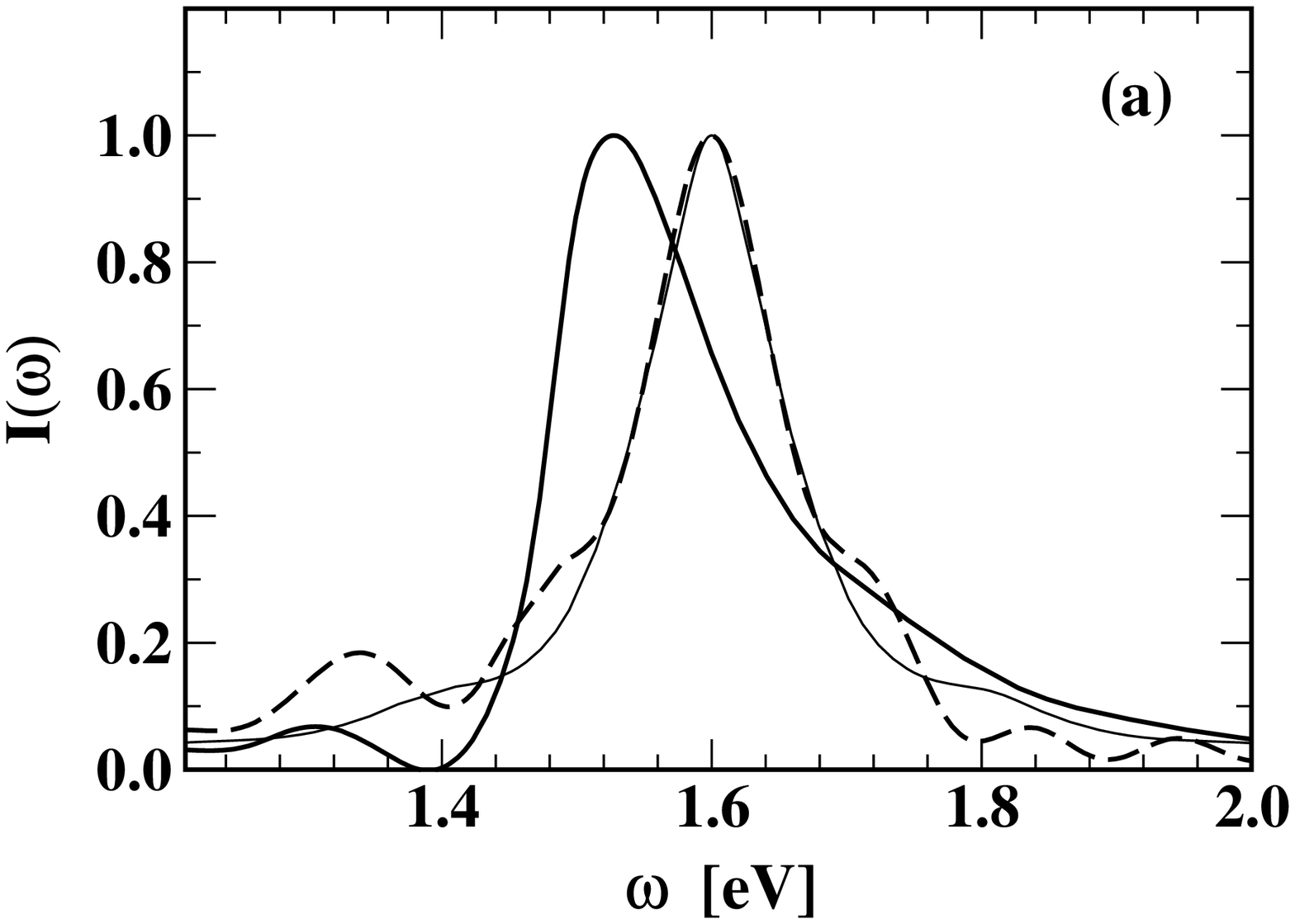}

    \vspace*{1ex}

      \hspace*{3ex}\includegraphics[clip,width=2.6in]{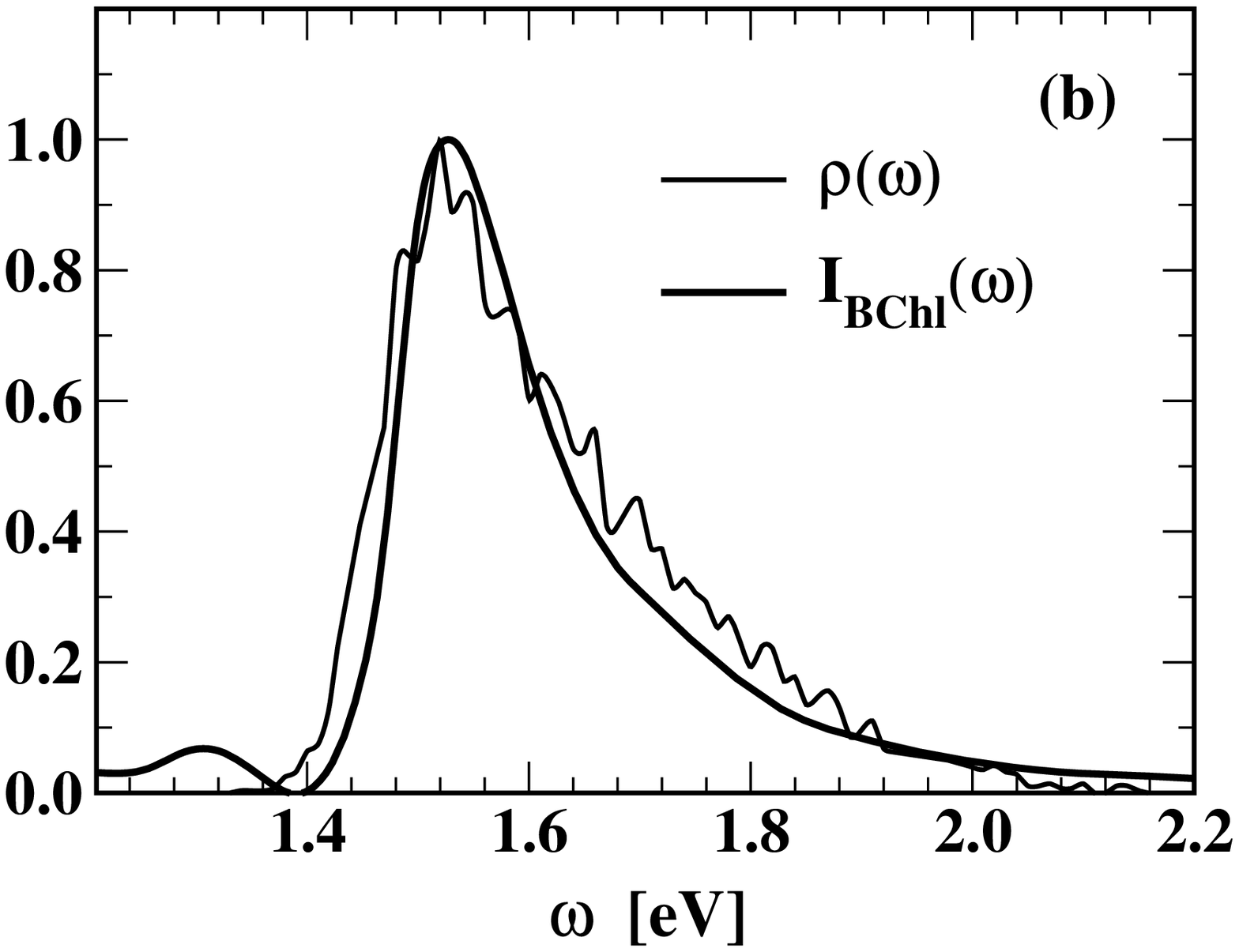}
      
    \vspace*{3ex}

    \caption{\textbf{(a)} Absorption spectra of individual BChls:
      $I(\omega)$ -- time series analysis (dashed curve), $I_0(\omega)$
      -- polaron model with ${\Phi''}_0=0$ (thin solid curve),
      $I_{BChl}(\omega)$ -- polaron model (thick solid curve).
      \textbf{(b)} Comparison between the normalized DOS $\rho(\omega)$
      and $I_{BChl}(\omega)$, obtained within the framework of the
      polaron model.}
    \label{fig:Iow}
  \end{center}
\end{figure}

\noindent In case of individual BChls, i.e., without excitonic coupling,
$F_k(t)=1$, and according to
Eqs.~(\ref{eq:Iw_},\ref{eq:Phi1a},\ref{eq:Phi2a}) the absorption
spectrum reads

\begin{mathletters}
\label{eq:IBChl+}
\begin{equation}
  \label{eq:IwBChl}
  I_{BChl}(\omega) \propto \int_0^\infty dt \exp[-{\Phi'}_0(t)]
  \cos[(\omega -\epsilon_0)t + {\Phi''}_0(t)]\;,
\end{equation}
where
\begin{equation}
  \label{eq:Phi10}
  {\Phi'}_0(t) = \int_0^\infty d\tau (t-\tau) {\cal C}(\tau)\;,
\end{equation}
and
\begin{equation}
  \label{eq:Phi20}
  {\Phi''}_0(t) = \int_0^\infty d\tau (t-\tau) {\cal D}_2(\tau)\;.
\end{equation}
\end{mathletters}
For the moment let us neglect the imaginary part of the cumulant
$\Phi_0(t)$ in Eq.~(\ref{eq:IwBChl}), i.e., we set ${\Phi''}_0(t)\approx
0$. It follows

\begin{equation}
  \label{eq:IwBChl0}
  I_{BChl}(\omega) \approx I_0(\omega) \propto \int_0^\infty dt
  \exp[-{\Phi'}_0(t)]  \cos[(\omega -\epsilon_0)t]\;,
\end{equation}
and one can easily see that $I_0(\omega)$ coincides (up to an
irrelevant normalization factor) with the cumulant approximation of
the line shape function $I(\omega)$, Eq.~(\ref{eq:abs_21}), derived
and employed in our computer simulation study. Therefore, it comes to
no surprise that in Fig.~\ref{fig:Iow}a the plots of the normalized
[with max$\{I(\omega)\} = 1$] line shape functions $I_0(\omega)$ and
$I(\omega)$ completely overlap in the peak region. The difference
between the curves away from the central peak is most likely due to
poor statistics and numerical artifacts introduced by the FFT
evaluation of the integrals for the time series analysis result
(\ref{eq:abs_21}). The peak of the absorption spectrum is located at
$\epsilon_0=1.6$~eV and the corresponding FWHM is 125~meV.

However, the actual line shape function $I_{BChl}(\omega)$, as given by
Eq.~(\ref{eq:IwBChl}), include a non-vanishing ${\Phi''}_0(t)$. The
corresponding result is also plotted in Fig.~\ref{fig:Iow}a. As one can
see, the contribution of ${\Phi''}_0(t)$ redshifts the peak of
$I_{BChl}(\omega)$ to 1.53~eV, and renders $I_{BChl}(\omega)$ asymmetric
with a somewhat larger FWHM of 154~meV. As illustrated in
Fig.~\ref{fig:Iow}b, $I_{BChl}(\omega)$ matches the corresponding DOS
$\rho(\omega)$ [see Eq.~(\ref{eq:ibm8}) and Fig.~\ref{fig:DOS}]. This is
not surprising since, as we have already mentioned, for a two level
system which is linearly coupled to a phonon bath, both the DOS and the
absorption spectrum are proportional to the imaginary part of the
corresponding Green's function \cite{mahan,nakajima}. The line shape
function (\ref{eq:abs_21}) of individual BChls derived within the
theoretical framework of Sec.~II does not account for the imaginary part
of the cumulant $\Phi_0(t)$ which suggests that the time development
operator $U(t,t_0)$ [Eq.~(\ref{eq:ddis_18b})] is not totally adequate
for calculating the absorption coefficient. The situation seems to be
similar to the commonly used \textit{stochastic models} in modeling the
absorption spectrum of excitonic systems coupled to a heat bath with a
finite time scale \cite{MUKA95}. In this case as well, the
autocorrelation function of the stochastic energy gap fluctuations is
assumed to be real (i.e., ${\Phi''}(t)$ is neglected), and as a result
the model violates the fluctuation dissipation theorem, and yields no
Stoke shift \cite{MUKA95}. This discrepancy is due to the fact that,
unlike in the case of the polaron model, the system-heat bath
interaction is not incorporated in a consistent manner.

\begin{figure}[t]
  \begin{center}
\vspace*{-4ex}
    \includegraphics[clip,width=3.4in]{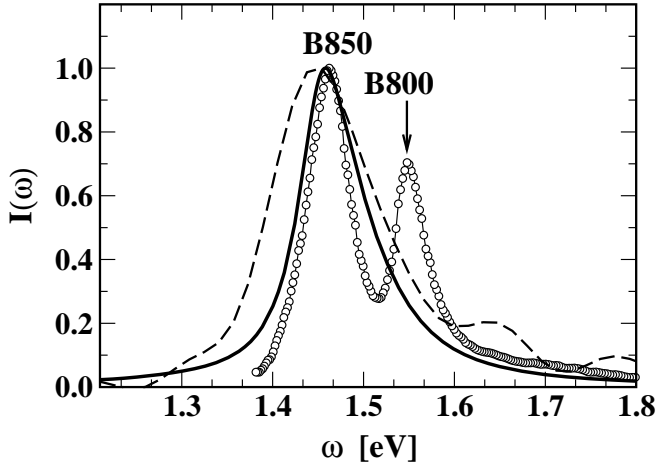}

    \vspace*{6ex}

    \caption{Normalized absorption spectra of the B850 BChls with
      excitonic coupling: $I_{B850}(\omega)$ -- polaron model (solid
      curve), $I_{TSA}(\omega)$ -- time series analysis (dashed curve)
      and $I_{exp}(\omega)$ -- experiment (circles).}
    \label{fig:Iw}
  \end{center}
\end{figure}

Within the framework of the polaron model, the absorption spectrum
$I_{B850}(\omega)$ of the excitonically coupled B850 BChls can be
calculated numerically by using
Eqs.~(\ref{eq:Iw_},\ref{eq:Phi12+},\ref{eq:D2},\ref{eq:w2J},\ref{eq:FF}).
The needed input quantities are: $T=300$~K, ${\cal C}(t)$, and
$\epsilon_k=\epsilon_{\pm 1}=1.52$~eV; these are derived from our
simulations and there are no free parameters in the above mentioned
equations. The computed $I_{B850}(\omega)$ is shown in
Fig.~\ref{fig:Iw}, along with the corresponding spectrum
$I_{TSA}(\omega)$ obtained through time series analysis,
[c.f.~Eqs.~(\ref{eq:abs_19},\ref{eq:abs_19a})], and the corresponding
experimental result $I_{exp}(\omega)$ \protect\cite{ZHAN2000}. The peak
of $I_{B850}$ is located at 1.46~eV ($\approx 849$~nm), and coincides
with the position of the other two spectra. $I_{B850}(\omega)$
(FWHM=88~meV) is somewhat broader than $I_{exp}(\omega)$ (FWHM=58~meV),
but is narrower than $I_{TSA}(\omega)$ (FWHM=138~meV).
The effect of exchange narrowing due to the excitonic coupling between
B850 BChls is manifest: FWHM is reduced from 154~meV
(Fig.~\ref{fig:Iw}) to 88~meV (Fig.~\ref{fig:Iow}a), which represents a
reduction in the width of the line shape function by a factor of 1.75.
This exchange narrowing is accounted for by the factor $F_k(t)$ in
Eq.~(\ref{eq:abs1}); since $|F_k(t)|\le 1$, this factor reduces the
value of the cumulant $\Phi_k(t)$, which leads to a narrowing of the
line shape function [recall that $\Phi_k=0$ result in a Dirac-delta
function type $I(\omega)$].
In contrast to the commonly used stochastic models \cite{MUKA95} for
describing the absorption spectrum of molecular aggregates, our polaron
model does not postulate the form of the autocorrelation function ${\cal
  C}(t)$, but rather uses this function as an input, derived from
computer simulations or from experiment. In this regard, our polaron
model provides a more realistic approach for evaluating optical
properties of molecular aggregates in general, and the B850 excitons in
particular. Indeed, a generic autocorrelation function ${\cal
  C}(t)=\Delta^2\exp(-\lambda t)$, assumed by most stochastic models,
would represent a gross oversimplification of the complicated structure
of ${\cal C}(t)$ as inferred from our computer simulations (see
Fig.~\ref{fig:corr-J}a). It is clear that this autocorrelation function
cannot be modeled by either a single or a sum of several exponentially
decaying functions, in spite of the fact that by a proper choice of the
values of the mean square fluctuations of the energy gap, $\Delta^2$,
and of the inverse relaxation time, $\lambda$, the stochastic model can
yield an almost perfect fit to the experimental spectrum. For example,
by assuming that in our case the broadening of the line shape function
is due to coupling to phonons with clearly separated high and low
frequency components, it can be shown that the real part of the cumulant
is approximately $\Phi'(t)\approx \Delta_{\ell}^2 t^2/2 + \Gamma_h t$,
where $\ell$ ($h$) refers to the low (high) frequency component and
$\Gamma_h=\Delta_h^2/\lambda_h$. (The imaginary part of the cumulant
$\Phi''(t)$ is assumed to affect only the shift of the absorption
spectrum, but not its broadening.) The first (second) term in $\Phi'(t)$
represents the short (long) time approximation of the cumulant brought
about by the low (high) frequency phonons. The corresponding line shape
function has a Voigt profile \cite{MUKA95}, i.e., is a convolution of a
Gaussian and Lorentzian due to low and high frequency phonons,
respectively. For $\Delta_s=25$~meV and $\Gamma_{\ell}=7$~meV, one can
obtain an almost perfect fit to the experimental B850 absorption
spectrum. However, in spite of this apparent agreement with experiment,
neither these numerical values nor the clear separation in the phonon
spectral density is justified in view of our computer simulation
results, and as such this kind of analysis should be avoided. In fact,
as it can be inferred from the plot of $J(\omega)$ shown in
Fig.~\ref{fig:corr-J}b, all phonon frequencies up to 0.24~eV contribute
to the cumulant $\Phi(t)$ and, therefore, to the absorption spectrum
$I(\omega)$. Thus, unlike stochastic models which employ fitting
parameters to account for the broadening of the experimental absorption
spectra, our polaron model incorporates dynamic disorder directly
through the phonon spectral function $J(\omega)$ obtained
from combined MD/quantum chemistry calculations, i.e., without any ad
hoc assumptions and fitting parameters. 

Further tests of our polaron model based on the calculation of other
optical properties of the system, such as the circular dichroism (CD)
spectrum, are in preparation.

\section{Discussion}

This paper presents a novel approach to the study of excitons in
light-harvesting complexes that combines molecular dynamics and
quantum chemistry calculations with time-dependent effective
Hamiltonian as well as polaron model type of analyses.

The molecular dynamics approach allowed us to observe, at the atomic
level, the dynamics of BChl motions and gain insight into the extent and
timescales of geometrical deformations of pigment and protein residues
at physiological temperatures. A comparison of the average distances and
angles between the B850 BChls to those found in the crystal structure
reveals an increased dimerization within the B850 ring, as compared to
the crystal structure. We observed diffusion of a water molecule into
the B800 BChl binding site for seven out of eight B800 BChls. The
average location of this water molecule agrees well with the crystal
structure location. Future quantum chemistry calculations will determine
whether this molecule plays a functional role. In this respect it is
interesting to note that the recently published structure of the
cyanobacterial photosystem I at 2.5~\AA resolution revealed also
ligations of Chla Mg ions involving side groups other than His, and, in
particular, water \cite{JORD2001}.

The time series analysis theory, based on the combined MD/QC
calculations resulted in an absorption spectrum that is about a factor
of two wider than the experimental spectrum. One of the reasons for this
discrepancy might be improper treatment, and subsequently an
overestimate, of the contribution of high-frequency modes within the
framework of our combined MD/QC calculations. We also note that the time
series analysis neglects some important quantum effects which may cause
the discrepancy between the computed and the experimental results. These
quantum effects are accounted for by the polaron model.  The amplitude
of fluctuations of the off-diagonal matrix elements, i.e., couplings
between BChls, was found to be at least two orders of magnitude smaller
than the corresponding fluctuation amplitude of the diagonal matrix
elements. We believe that in spite of the possible overestimate of the
fluctuation associated with the high frequency modes, we can safely
conclude that the disorder in the B850 system is diagonal rather than
off-diagonal in nature.

The time-dependent effective Hamiltonian description revealed an exciton
spectrum red-shifted as compared to the spectrum of individual BChls.
This red-shift is well known in J-aggregates, and is attributed to the
transferring of the dipole strength into low-energy exciton states.

The observation that the fluctuations of the BChl site energies could be
attributed largely to high frequency intramolecular vibrational modes,
with energy $\omega_0\sim 1,600-1,700$~cm$^{-1}$, prompted us to model
the effect of dynamic disorder on the B850 excitons by employing a
polaron Hamiltonian, which describes both excitons and the coupled
single phonon mode by quantum mechanics. The strength of the
exciton-phonon coupling $g$ is related to the RMSD of the fluctuating
site energies, and was found to be weak.  By employing standard
perturbation theory we investigated the effect of dynamic disorder
(i.e., exciton-phonon coupling) on the energy dispersion and
localization of the excitons. We found that in contrast to static
disorder, which leads to a broadening of the exciton bandwidth, dynamic
disorder due to high frequency vibrational modes leads to a reduction of
the width of the exciton energy band.  Also, our polaron calculations
showed that dynamic disorder reduces only slightly the exciton
delocalization length (from around 6 BChls to 5 BChls at room
temperature), confirming previous results according to which the main
mechanism responsible for exciton localization in LH-II rings is thermal
averaging \cite{RAY99}.

By employing the cumulant expansion method, we calculated, in the
framework of the polaron model, the absorption spectrum of the B850
BChl, with and without exciton coupling, for a single phonon mode as
well as for a distribution of phonons. The absorption spectrum of the
excitonic system coupled to a single high frequency phonon $\omega_0$
(intramolecular vibronic mode) is given by a series of Dirac-delta
functions (stick spectrum), with different weights. In order to obtain a
realistic, broadened absorption spectrum, which can be compared with the
one measured experimentally we had to included the effect of the entire
phonon distribution through the phonon spectral density $J(\omega)$. We
have shown that $J(\omega)$, which accounts for both coupling strengths
and frequencies of the individual phonon components, can be determined
from the autocorrelation function ${\cal C}(t)$ of the energy gap
fluctuations of individual BChls, readily available from our combined
MD/quantum chemistry simulation. We found that in our model only three
inputs, namely autocorrelation function ${\cal C}(t)$, temperature $T$
and optically active exciton level $\epsilon_k$, are needed to calculate
the absorption spectrum. Considering the fact that these inputs come
directly from our MD/QC calculations, and that we have no free
parameters, the agreement between the spectrum obtained from the polaron
model calculations and the experimental spectrum is remarkable.

\acknowledgements

The authors thank Nicolas Foloppe for communication of data and
advice on force-field parameterizations, Jerome Baudry for advice
on force-field parameterizations, Justin Gullingsrud and James
Phillips for much help in getting started with the molecular
dynamics simulations, Lubos Mitas and Sudhakar Pamidighantam for
 advice on quantum chemistry simulations and on
running those on NCSA clusters, and Shigehiko Hayashi and Emad
Tajkhorshid for advice on quantum chemistry absorption spectra
calculations. AD thanks the Fleming group for useful discussions. 
This work was supported by grants from the National
Science Foundation (NSF BIR 94-23827 EQ and NSF BIR-9318159), the
National Institutes of Health (NIH PHS 5 P41 RR05969-04), the
Roy J.~Carver Charitable Trust, and the MCA93S028 computertime grant.


\end{document}